\documentclass{emulateapj}

\slugcomment{ver.Jun.18; draft 1 pages}
\slugcomment{ver.Aug.1; draft 20 pages}
\slugcomment{ver.Aug.17; draft 31 pages}
\slugcomment{ver.Sep.16; draft 36 pages}
\slugcomment{ver.Sep.30; draft 43 pages}
\slugcomment{ver.1.2: 2010.10.05}
\slugcomment{ver.2.3: 2010.10.24}
\slugcomment{ver.2.4: 2010.10.31}
\slugcomment{ver.3: 2011.01.13; revisions after the referee's comments}
\slugcomment{ver.3: 2011.02.04; revisions after the referee's comments}
\slugcomment{ver.3.1: 2011.02.16; revisions after the referee's comments}
\slugcomment{ver.3.2: 2011.02.24; revisions after the referee's comments}
\slugcomment{ver.4: 2011.04.11; revisions after the referee's comments-2}
\slugcomment{Received 2010 November 3; accepted 2011 April 11}

\shorttitle{Ly$\alpha$ emitters at $z=6.5$}
\shortauthors{Kashikawa et al.}

\begin{document}


\title{Completing the Census of Ly$\alpha$ Emitters at the Reionization Epoch\altaffilmark{1,2}}

\author{
Nobunari Kashikawa\altaffilmark{3,4} 
Kazuhiro Shimasaku\altaffilmark{5,6}, 
Yuichi Matsuda\altaffilmark{7},
Eiichi Egami\altaffilmark{8},
Linhua Jiang\altaffilmark{8},
Tohru Nagao\altaffilmark{9}, 
Masami Ouchi\altaffilmark{10}, 
Matthew A. Malkan\altaffilmark{11},
Takashi Hattori\altaffilmark{12}, 
Kazuaki Ota\altaffilmark{10}, 
Yoshiaki Taniguchi\altaffilmark{9},\\
Sadanori Okamura\altaffilmark{5,6}, 
Chun Ly\altaffilmark{13},
Masanori Iye\altaffilmark{3,4}, 
Hisanori Furusawa\altaffilmark{14}, 
Yasuhiro Shioya\altaffilmark{9},\\
Takatoshi Shibuya\altaffilmark{4},
Yoshifumi Ishizaki\altaffilmark{4},
 and
Jun Toshikawa\altaffilmark{4}
%
%
%
}




\email{n.kashikawa@nao.ac.jp}


\altaffiltext{1}{The data presented herein were partly obtained at the W. M. Keck Observatory, which is operated as a scientific partnership among the California Institute of Technology, the University of California, and the National Aeronautics and Space Administration. 
The Observatory was made possible by the generous financial support of the W. M. Keck Foundation.}
\altaffiltext{2}{Based in part on data collected at the Subaru Telescope, which is operated by the National Astronomical Observatory of Japan.}
\altaffiltext{3}{Optical and Infrared Astronomy Division, National Astronomical Observatory, Mitaka, Tokyo 181-8588, Japan; n.kashikawa@nao.ac.jp}
\altaffiltext{4}{Department of Astronomy, School of Science, Graduate University for Advanced Studies, Mitaka, Tokyo 181-8588, Japan}
\altaffiltext{5}{Department of Astronomy, University of Tokyo, Hongo, Tokyo 113-0033, Japan}
\altaffiltext{6}{Research Center for the Early Universe, University of Tokyo, Hongo, Tokyo 113-0033, Japan}
\altaffiltext{7}{Department of Physics, Durham University, South Road, Durham DH1 3LE, UK}
\altaffiltext{8}{Steward Observatory, University of Arizona, 933 North Cherry Avenue, Tucson, AZ 85721, USA}
\altaffiltext{9}{Research Center for Space and Cosmic Evolution, Ehime University, Bunkyo-cho, Matsuyama 790-8577, Japan}
\altaffiltext{10}{Institute for Cosmic Ray Research, University of Tokyo, 5-1-5 Kashiwa-no-Ha, Kashiwa City, Chiba 77-8582, Japan}
\altaffiltext{11}{Department of Physics and Astronomy, University of California, Los Angeles, CA 90095-1547, USA}
\altaffiltext{12}{Subaru Telescope, National Astronomical Observatory of Japan, 650 North A'ohoku Place, Hilo, HI 96720, USA}
\altaffiltext{13}{Space Telescope Science Institute, 3700 San Martin Drive, Baltimore, MD 21218, USA}
\altaffiltext{14}{Astronomy Data Center, National Astronomical Observatory, Mitaka, Tokyo 181-8588, Japan}


\begin{abstract}
We carried out extended spectroscopic confirmations of Ly$\alpha$ emitters (LAEs) at $z=6.5$ and $5.7$ in the Subaru Deep Field.
Now, the total number of spectroscopically confirmed LAEs is $45$ and $54$ at $z=6.5$ and $5.7$, respectively, and at least $81\%$ ($70\%$) of our photometric candidates at $z=6.5$ ($5.7$) have been spectroscopically identified as real LAEs.
We made careful measurements of the Ly$\alpha$ luminosity, both photometrically and spectroscopically, to accurately determine the Ly$\alpha$ and rest-UV luminosity functions (LFs). 
The substantially improved evaluation of the Ly$\alpha$ LF at $z=6.5$ shows an apparent deficit from $z=5.7$ at least at the bright end, and a possible decline even at the faint end, though small uncertainties remain.
The rest-UV LFs at $z=6.5$ and $5.7$ are in good agreement, at least at the bright end, in clear contrast to the differences seen in the Ly$\alpha$ LF.
These results imply an increase in the neutral fraction of the intergalactic medium from $z=5.7$ to $6.5$.
The rest-frame equivalent width (EW$_0$) distribution at $z=6.5$ seems to be systematically smaller than $z=5.7$, and it shows an extended tail toward larger EW$_0$.
The bright end of the rest-UV LF can be reproduced from the observed Ly$\alpha$ LF and a reasonable EW$_0-$UV luminosity relation.
Integrating this rest-UV LF provides the first measurement of the contribution of LAEs to the photon budget required for reionization.
The derived UV LF suggests that the fractional contribution of LAEs to the photon budget among Lyman break galaxies significantly increases toward faint magnitudes.
Low-luminosity LAEs could dominate the ionizing photon budget, though this inference depends strongly on the uncertain faint-end slope of the Ly$\alpha$ LF.

\end{abstract}






\section{Introduction}

Cosmic reionization was a major event in the early history of the universe.
It is a drastic phase transition of the intergalactic medium (IGM) and is closely related to the birth of the first galaxies; however, it is still unclear when and how reionization occurred.
The polarization measurement of the cosmic microwave background (CMB) by Wilkinson Microwave Anisotropy Probe (WMAP) implies reionization at $z=10.9\pm1.4$ \citep{kom09}, and the complete Gunn-Peterson (GP) trough of the Sloan Digital Sky Survey (SDSS) QSOs suggests that cosmic reionization ended at $z\sim 6$ (e.g., \citealp{fan06}).
In addition to these unknowns, it is unclear which objects were responsible for the cosmic reionization.
Although QSOs are expected to be the main contributor at the bright end of the luminosity function (LF) of ionizing sources, the QSO population alone cannot account for all the required ionizing photons (\citealp{jia08}, \citealp{wil05}).
Star-forming galaxies such as Lyman break galaxies (LBGs) and Ly$\alpha$ emitters (LAEs) at the reionization epoch are the only alternatives that could dominate the LF at the faint end.

The LAEs are one class of high-$z$ star-forming galaxies. 
Detecting their strong Ly$\alpha$ emissions is feasible even beyond $z=6$ using deep narrow band (NB) imaging. 
Along with LBGs, they tell us about early star formation history and initial structure formation. 
In addition, they may serve as valuable observational tools for probing the cosmic reionization process. 
Ly$\alpha$ photons are sensitive to the physical state of the IGM. 
It is expected that the damping wing from the surrounding neutral IGM attenuates Ly$\alpha$ photons so significantly that the Ly$\alpha$ emission flux will be reduced; therefore, it is naturally expected that the Ly$\alpha$ LF of LAEs should decline as it traces earlier times in the reionization epoch \citep{hai99, mal04}.
The LAE population appears to have a similar number density over a long-time period from $z=3$ to $6$ (e.g., \citealp{ouc08}).
Thus, if a sharp decline appears in the Ly$\alpha$ LF of LAEs, it could result from attenuation by the neutral IGM.
Consequently, the observed census of LAEs during the reionization period could trace the neutral fraction of IGM hydrogen, $x_{\rm H I}$.
The census of observable LAEs beyond $z=6$ is sensitive to $x_{\rm H I}$ of the universe.

This ^^ ^^ Ly$\alpha$ test" has an advantage in that it is sensitive even at $x_{\rm HI}> 10^{-3}$, which is the upper limit for the application of the GP test.
In addition, this method uses galaxies that are more abundant than QSOs or gamma-ray bursts (GRBs), so it will yield a volume-averaged estimate for the neutral fraction. 
Significant variation in IGM transmissions among different QSO lines-of-sight \citep{djo06} suggests that the reionization process is spatially patchy.
In the future, it will be possible to investigate field-to-field variation in the neutral fraction, providing qualitative estimates of this spatially patchy reionization process.
Alternatively, the change in the Ly$\alpha$ LF could be caused by galaxy evolution.
Distinguishing IGM attenuation from galaxy evolution is crucial for this test.
However, recent systematic surveys for LAEs at lower-$z$ revealed that the Ly$\alpha$ LF exhibits almost no evolution from $z=3$ to $z=6$, though the physical reason for this is unclear.
In any case, this result strongly supports the viability of the Ly$\alpha$ test, in which any decline of such static LAE LF at $z>5.7$ would be caused by an IGM attenuation.
Nevertheless, a part of the LF decline should be caused by the intrinsic galaxy evolution of LAEs; therefore, it is important to compare the rest-UV LFs based on the LAE sample at the same time as the Ly$\alpha$ test is made, because the rest-UV continuum flux is not sensitive to the neutral IGM.

Several model predictions of the Ly$\alpha$ LF at the reionization epoch agree that the amplitude of the LF decreases as $x_{\rm HI}$ increases (\citealp{mcq07}; \citealp{hai05}; \citealp{led05}; \citealp{dij07b}; \citealp{mes07}; \citealp{kob07}; \citealp{ili08}; \citealp{day09}).
The H {\sc ii} bubbles of bright LAEs clustered in the overdense regions would effectively overlap, creating a larger H {\sc ii} bubble with a high ionization fraction, which would significantly reduce the Ly$\alpha$ flux attenuation.
As a result, the bright LAEs that are expected to reside in a high-density environment should be readily observed, whereas faint LAEs are more severely attenuated.
This results in a smooth decrease in the amplitude of the LF toward higher $x_{\rm HI}$.
To compare with these model predictions, we have to observe a large sample of high-$z$ LAEs.

In our previous papers (\citealp{kas06}: K06, \citealp{shi06}: S06), we have presented spectroscopic confirmations of $17$ and $34$ LAEs at $z=6.5$ and $5.7$, respectively, in the Subaru Deep Field (SDF); see also \citet{kod03} and \citet{tan05} for LAEs at $z=6.5$.
The sample consisted of objects with excess flux in narrow-band $NB921$ ($\lambda_c=9196$ \AA, FWHM=$132$ \AA) and $NB816$ ($\lambda_c=8150$ \AA, FWHM=$120$ \AA) images.
We found that the Ly$\alpha$ LF at $z=6.5$ reveals an apparent deficit at least at the bright end, corresponding to $\sim0.75$ magnitudes fainter, compared with that observed at $z=5.7$.
The decline in the Ly$\alpha$ LF from $z=5.7$ to $6.5$ could imply a substantial transition in the IGM ionizing state between these epochs, suggesting that reionization was not complete at $z=6.5$.
The trend was also confirmed in the LAE sample at $z\approx7$ \citep{iye06}.
The Ly$\alpha$ luminosity density of LAEs did not change from $z=3$ to $z=5.7$, and it gradually decreased from $z=5.7$ to $7$ \citep{ota08}.
The abrupt drop in luminosity density from $z=5.7$ to $7.0$ did not seem to be caused by continuous galaxy evolution.
This gradual decline recalls the IGM attenuation, though we have few spectroscopically identified galaxies at $z=7$.

Turning to the faint end of the LF below $L$(Ly$\alpha$)$=5\times10^{42}$, the amplitude difference between our photometrically- and spectroscopically-determined estimates was too large to constrain the faint end.
Our faint spectroscopic sample at both $z=6.5$ and $5.7$ was still so small that we could not reliably identify a significant difference in the LF between these two epochs at the faint end.
We neither determined the faint-end slope of the LF nor constrained the true contribution of LAEs to the entire photon budget required for full reionization.

\begin{deluxetable*}{llllllll}
\tabletypesize{\footnotesize}
\tablecaption{Summary of Spectroscopic Observations\label{tab_obs}}
\tablewidth{0pt}
\tablehead{
\colhead{Observational run} & \colhead{Instrument} & \colhead{date(UT)} & \colhead{T$_{\rm integ}$(ksec)} & \colhead{seeing(arcsec)} & \colhead{N$_{\rm mask}$\tablenotemark{a}} & \colhead{N$_{65}$\tablenotemark{b}} & \colhead{N$_{57}$\tablenotemark{c}}
}
\startdata
2006\tablenotemark{d} & FOCAS  & Apr.26        & 7.2---9.0   & 0.5---1.0 & 2 & 2  & 0  \\
2007                  & FOCAS  & May19---22,24 & 10.8---16.2 & 0.5---1.2 & 5 & 15 & 4  \\
2008                  & DEIMOS & Apr.30, May1  & 5.4---10.8  & 0.7---1.1 & 4 & 1  & 2  \\
2009                  & DEIMOS & Apr.26---27   & 10.8---16.2 & 0.7---0.8 & 3 & 8  & 10 \\
2009\tablenotemark{e} & DEIMOS & Apr.24---25   & 10.8---12.6 & 0.7---0.8 & 2 & 2  & 1  \\
2010\tablenotemark{f} & FOCAS  & Mar.18---21   & 18.0        &           & 2 & 0  & 3  \\
Total                 &        &               &             &           &   & 28 & 20 \\
\enddata
\tablenotetext{a}{The total number of masks.}
\tablenotetext{b}{The total number of identified LAEs at $z=6.5$.}
\tablenotetext{c}{The total number of identified LAEs at $z=5.7$.}
\tablenotetext{d}{These were identified in the course of other observational program in Nagao et al. (2007).}
\tablenotetext{e}{These were identified in the course of other observational program in L. Jiang et al. (2011, in preparation).}
\tablenotetext{f}{These were identified in the course of other observational program in M. Iye et al. (in preparation).}
\end{deluxetable*}

\epsscale{1.2}
\begin{figure}
\plotone{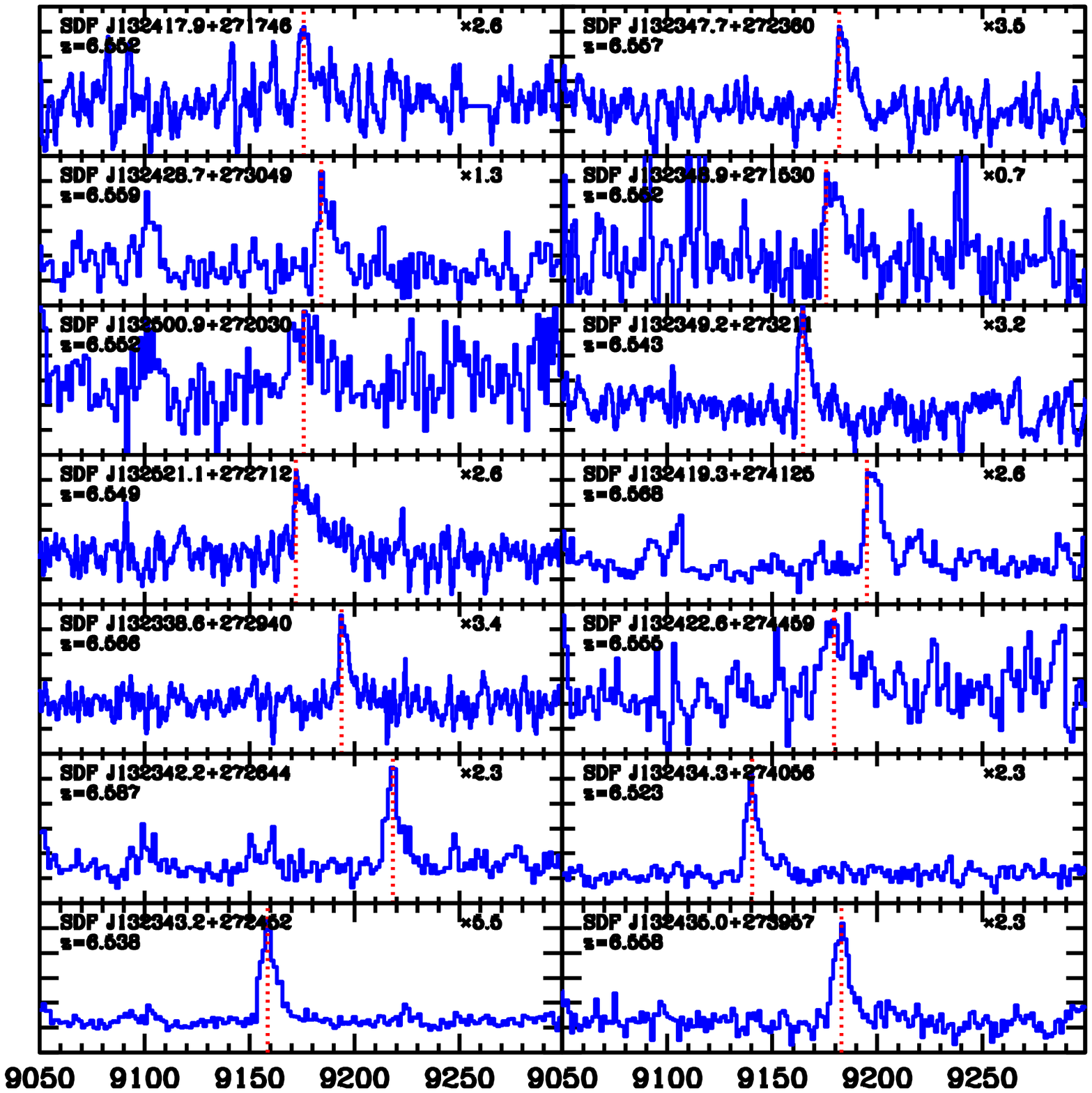}
\plotone{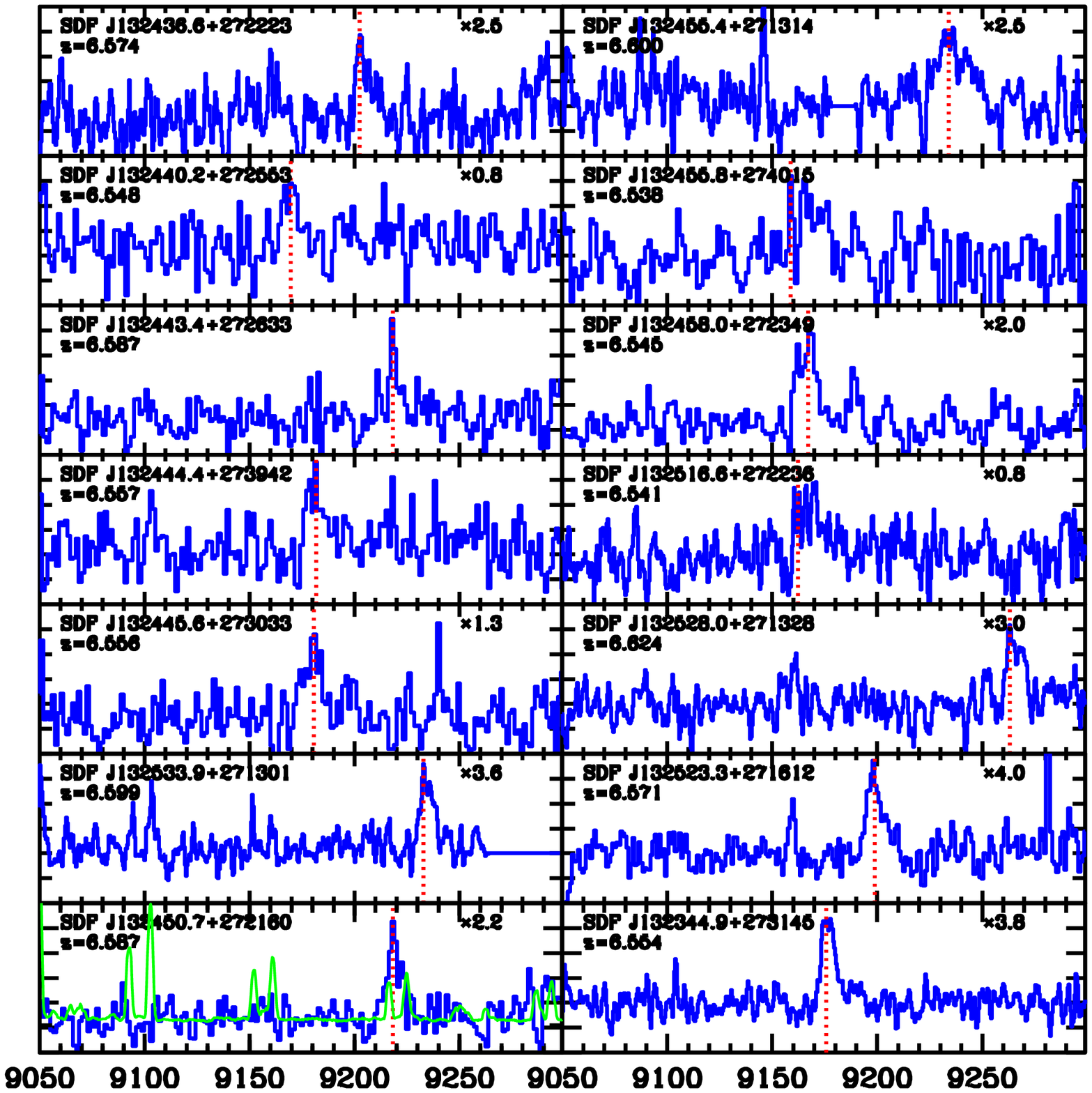}
\epsscale{1.0}
\caption{
Spectra of $28$ spectroscopically confirmed LAEs at $z=6.5$.
ID number and redshift are shown in the upper left corner of each panel.
Spectra taken by DEIMOS were smoothed with a 3-pixel boxcar.
The vertical dotted line indicates the center of the Ly$\alpha$ emission line.
The vertical scale is marked in $0.1\times10^{-18}$ erg s$^{-1}$ cm$^{-2}$ \AA$^{-1}$, and a scaling factor to obtain the correct scale appears in the right corner in each panel.
The sky spectrum is overplotted on the left bottom panel with an arbitrary flux scale.
\label{fig_spe65}}
\end{figure}

\epsscale{1.2}
\begin{figure}
\plotone{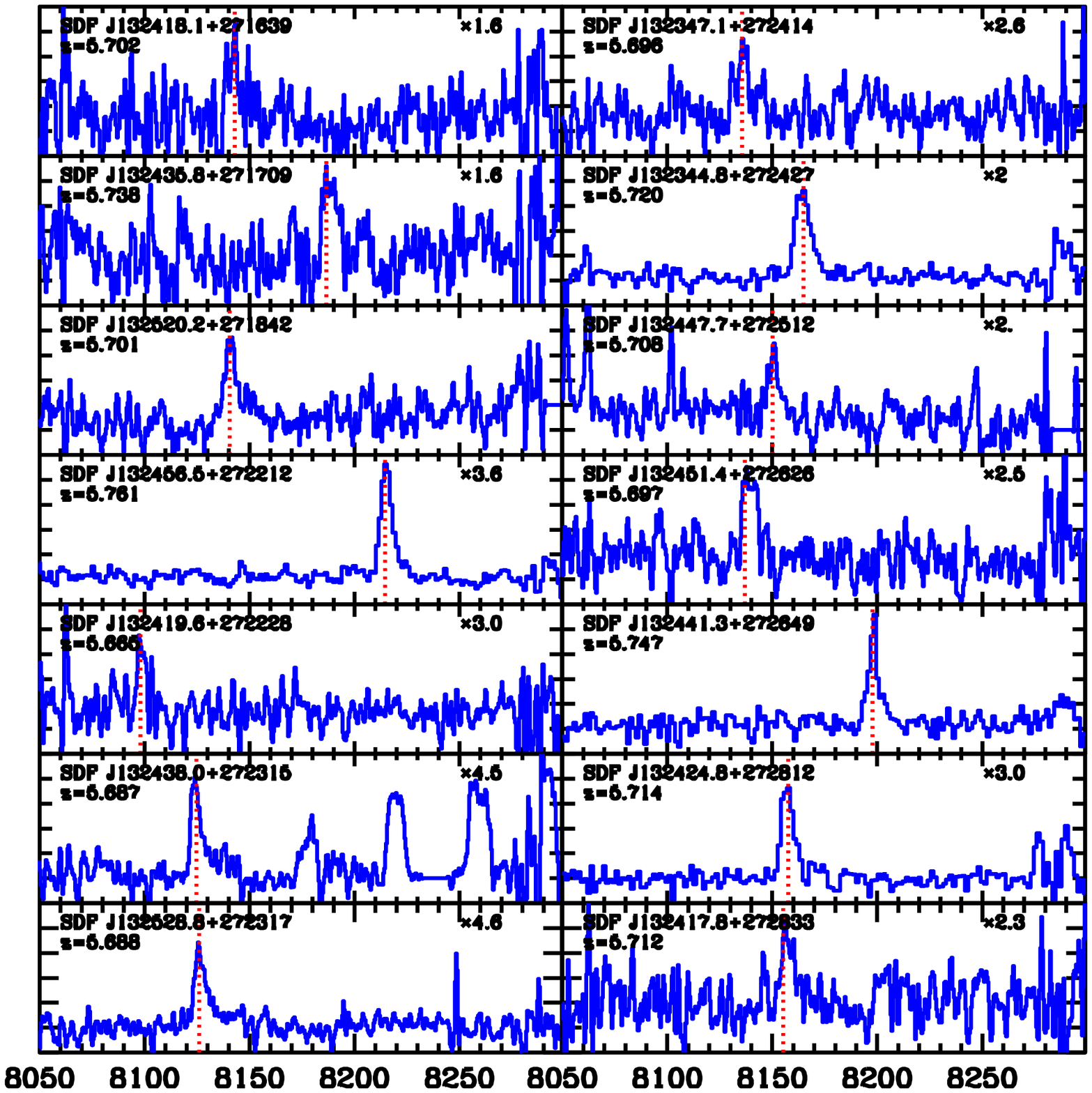}
\plotone{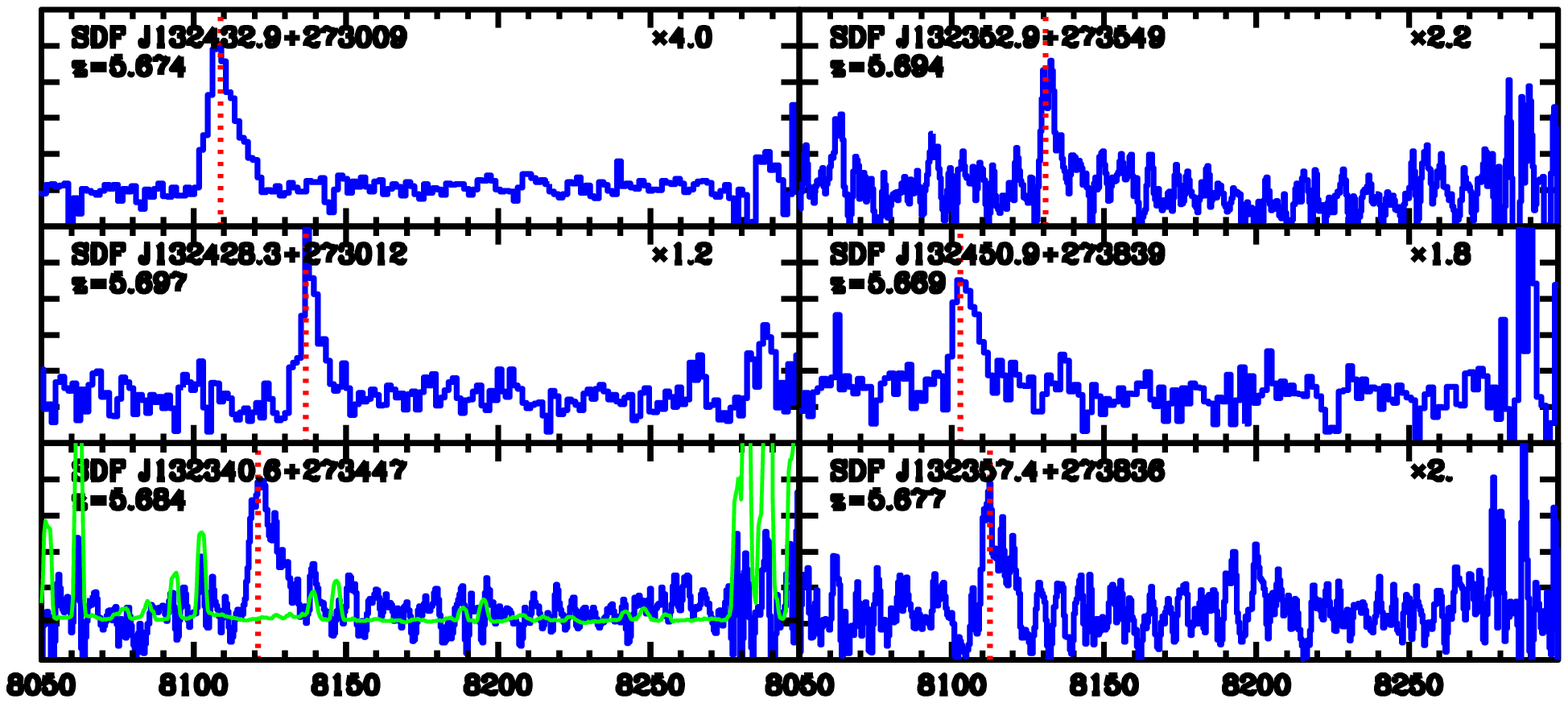}
\vspace*{-5.0cm}
\epsscale{1.0}
\caption{
Same as Figure~\ref{fig_spe65}, but for $20$ LAEs at $z=5.7$.
\label{fig_spe57}}
\end{figure}

In this paper, we describe our extended spectroscopic confirmations of LAEs at $z=6.5$ and $5.7$ after K06 and S06 to determine more accurately the faint end of the LAE LFs close to the reionization epoch.
Spectroscopic confirmations enhance the reliability of the LAE sample, eliminating possible contaminations from emission-line galaxies at lower $z$, and improving the derivation of the Ly$\alpha$ luminosity with precise redshifts.
A large number of spectroscopically identified LAEs enable us to construct more accurate LFs in both the Ly$\alpha$ luminosity, which is sensitive to the neutral IGM, and the rest-UV luminosity, which is insensitive to it.
The larger spectroscopic sample also provides better statistical estimates of the contamination rate, increasing the reliability of our photometric LAE sample.
Purely photometric detection of LAE candidates might have non-negligible ambiguities in estimating both Ly$\alpha$ luminosities, which is sensitive to the neutral IGM, and the rest-UV luminosity, which is insensitive to it, especially at high-$z$ beyond $z=6.5$, where it is difficult to detect them even with broadband imaging (\citealp{hib10}; \citealp{til10}; \citealp{ota10}).
An increase in the spectroscopic sample size could also improve the stacked spectrum, which may provide insights into either the internal dynamics of the LAE or the IGM properties (K06).

The Subaru Suprime-Cam has a very large field of view (FOV; $27\times34$ arcmin$^2$).
Wide-field imaging increases the chance of discovering rare objects, such as the most distant galaxies \citep{iye06}.
Furthermore, it is generally predicted in almost all models of the reionization process that an overlapping H {\sc ii} bubble at the end of the reionization could be as large as $10$ physical Mpc \citep{bar04, wyi04, fur04}, which corresponds to the FOV size of the Subaru Suprime-Cam.
Thus, wide FOV observations are essentially required to achieve an unbiased picture of the universe at the reionization period.
Our LAE samples were obtained from a general blank field without resorting to amplification of gravitational lensing by foreground clusters, providing reliable statistics about their LFs.
High-$z$ surveys rendered by gravitational lensing, which are accessible to low-luminosity sources (\citealp{ric08}, \citealp{sta07}, \citealp{hu02}), are complementary to our survey.

This paper is organized as follows:
In Section 2, we review our photometric LAE sample and initial spectroscopic identifications, presented in our previous studies.
In Section 3, we describe our new spectroscopically identified LAE sample at $z=6.5$ and $5.7$.
In Section 4, we accurately derive the Ly$\alpha$ and rest-UV luminosities of our LAE sample, comparing the Ly$\alpha$ and UV LFs between $z=6.5$ and $5.7$.
Comparisons of the rest-frame equivalent width (EW$_0$) and stacked Ly$\alpha$ emission profile are presented in Section 5 and Section 6.
In Section 7, we present a new method for deriving the rest-UV LF of LAEs, using the Ly$\alpha$ LF and EW$_0$ distribution, to constrain the photon budget required for reionization.
A summary of the paper is provided in Section 8, with some discussion of the implications for reionization based on our results.

Throughout the paper, we assume cosmology parameters: $\Omega_{\rm m}=0.3$, $\Omega_\Lambda=0.7$, and $H_0=70$ $h_{70}$ km s$^{-1}$ Mpc$^{-1}$. 
These parameters are consistent with recent CMB constraints \citep{kom09}.
Magnitudes are given in the AB system.

\begin{deluxetable*}{llccccrcrc}
\tabletypesize{\scriptsize}
\tablecaption{Spectroscopic Properties of $z=6.5$ LAEs \label{tab_laespec65}}
\tablewidth{0pt}
\tablehead{
\colhead{ID\tablenotemark{a}} & \colhead{NAME} & \colhead{z\tablenotemark{b}} & \colhead{$f^{\rm spec}$(Ly$\alpha$)\tablenotemark{c}} & \colhead{$L^{\rm spec}$(Ly$\alpha$)\tablenotemark{d}} & \colhead{FWHM\tablenotemark{e}} & \colhead{$S_w$} & \colhead{OBS.\tablenotemark{f}}\\
   & & & ($10^{-18}$ ergs s$^{-1}$ cm$^{-2}$) & ($10^{42} h_{70}^{-2}$ ergs s$^{-1}$) & (\AA) & \multicolumn{1}{c}{(\AA)} & \\
}
\startdata

12 &    SDF J132417.9+271745 &  6.552 & 10.5 & 5.14 & 9.05 & $9.33  \pm  2.38$  & D\\ 
13 &    SDF J132428.7+273049 &  6.559 & 5.35 & 2.62 & 10.1 & $7.49  \pm  1.72$  & F\\ 
14 &    SDF J132500.9+272030 &  6.552 & 8.78 & 4.29 & 19.2 & $8.45  \pm  3.12$  & F\\ 
19 &    SDF J132521.1+272712 &  6.549 & 9.56 & 4.67 & 12.4 & $12.91 \pm  3.17$  & D\\ 
22 &    SDF J132338.6+272940 &  6.566 & 9.59 & 4.71 & 5.04 & $12.0  \pm  3.14$  & D\\ 
23 &    SDF J132342.2+272644 &  6.587 & 12.5 & 6.19 & 6.50 & $6.17  \pm  0.96$  & F\\ 
24 &    SDF J132343.2+272452 &  6.538 & 21.5 & 10.4 & 9.10 & $8.74  \pm  0.56$  & F\\ 
25 &    SDF J132347.7+272360 &  6.557 & 10.2 & 4.99 & 9.53 & $7.41  \pm  1.84$  & D\\ 
26 &    SDF J132348.9+271530 &  6.552 & 4.66 & 2.28 & 16.0 & $16.66 \pm  4.56$  & F\\ 
27 &    SDF J132349.2+273211 &  6.543 & 9.62 & 4.69 & 7.99 & $8.16  \pm  4.83$  & D\\ 
34 &    SDF J132419.3+274125 &  6.568 & 11.9 & 5.85 & 9.41 & $3.54  \pm  0.59$  & F\\ 
35 &    SDF J132422.6+274459 &  6.555 & 7.96 & 3.90 & 18.5 & $11.30 \pm  4.73$  & F\\ 
40 &    SDF J132434.3+274056 &  6.523 & 9.27 & 4.49 & 7.45 & $8.89  \pm  1.10$  & F\\ 
41 &    SDF J132435.0+273957 &  6.558 & 12.5 & 6.10 & 8.83 & $5.25  \pm  1.40$  & F\\ 
42 &    SDF J132436.5+272223 &  6.574 & 6.37 & 3.14 & 9.39 & $9.07  \pm  2.88$  & D\\ 
43 &    SDF J132440.2+272553 &  6.548 & 6.10 & 2.98 & 16.5 & $4.03  \pm  3.61$  & F\\ 
44 &    SDF J132443.4+272633 &  6.587 & 6.85 & 3.39 & 16.0 & $6.89  \pm  2.14$  & F\\ 
45 &    SDF J132444.4+273942 &  6.557 & 6.97 & 3.42 & 14.0 & $14.60 \pm  22.02$ & F\\ 
46 &    SDF J132445.6+273033 &  6.556 & 10.0 & 4.91 & 23.2 & $16.62 \pm  4.95$  & F\\ 
47\tablenotemark{g} & SDF J132533.9+271301 &  6.599 & 6.53 & 5.94 & 8.84 & $6.62 \pm 0.51$ & D\\ 
49 &    SDF J132450.7+272160 &  6.587 & 10.8 & 5.35 & 9.14 & $3.98  \pm  1.22$  & F\\ 
50 &    SDF J132455.4+271314 &  6.660 & 13.5 & 6.70 & 18.1 & $6.04  \pm  7.09$  & D\\ 
51 &    SDF J132455.8+274015 &  6.538 & 7.06 & 3.44 & 25.2 & $10.55 \pm  6.75$  & F\\ 
53 &    SDF J132458.0+272349 &  6.545 & 17.7 & 8.65 & 17.3 & $3.99  \pm  2.72$  & F\\ 
56 &    SDF J132516.6+272236 &  6.541 & 6.59 & 3.21 & 14.5 & $6.60  \pm  13.6$  & D\\ 
57 &    SDF J132528.0+271328 &  6.624 & 17.5 & 8.79 & 10.1 & $6.41  \pm  4.03$  & D\\ 
60 &    SDF J132523.3+271612 &  6.571 & 15.9 & 7.82 & 10.9 & $8.87  \pm  0.59$  & F\\ 
61 &    SDF J132344.9+273145 &  6.554 & 14.5 & 7.08 & 8.04 & $7.89  \pm  3.75$  & D\\ 

\enddata
\tablenotetext{a}{The object IDs are those of T05, except ID=60 and 61, that are not listed in the photometric catalog of T05.}
\tablenotetext{b}{The redshift was derived from the wavelength of the flux peak in an observed spectrum assuming the rest wavelength of Ly$\alpha$ to be $1215$\AA.
These measurements could be overestimated in the case of significant damping wings by IGM.
Also, the observed peak position was slightly shifted redward due to instrumental resolution. See Figure \ref{fig_prof}.}
\tablenotetext{c}{The observed line flux corresponds to the total amount of the flux within the line profile. The slit-loss was corrected.}
\tablenotetext{d}{No dust absorption correction was applied.}
\tablenotetext{e}{Corrected for instrumental broadening.}
\tablenotetext{f}{Observed with FOCAS (F) or DEIMOS (D).}
\tablenotetext{g}{The coordinates of T05 ID-47 were incorrectly appeared in T05.}

\end{deluxetable*}

\begin{deluxetable*}{llccccrcrc}
\tabletypesize{\scriptsize}
\tablecaption{Spectroscopic Properties of $z=5.7$ LAEs \label{tab_laespec57}}
\tablewidth{0pt}
\tablehead{
\colhead{ID\tablenotemark{a}} & \colhead{NAME} & \colhead{z\tablenotemark{b}} & \colhead{$f^{\rm spec}$(Ly$\alpha$)\tablenotemark{c}} & \colhead{$L^{\rm spec}$(Ly$\alpha$)\tablenotemark{d}} & \colhead{FWHM\tablenotemark{e}} & \colhead{$S_w$} & \colhead{OBS.\tablenotemark{f}}\\
   & & & ($10^{-18}$ ergs s$^{-1}$ cm$^{-2}$) & ($10^{42} h_{70}^{-2}$ ergs s$^{-1}$) & (\AA) & \multicolumn{1}{c}{(\AA)} & \\
}
\startdata

34549  & SDF J132418.1+271639 & 5.702 & 9.56 & 3.38 & 16.9 & $9.97  \pm  1.99$  & D\\
36849  & SDF J132435.8+271709 & 5.738 & 6.92 & 2.49 & 12.0 & $32.8  \pm  11.7$  & D\\
45159  & SDF J132520.2+271842 & 5.701 & 10.5 & 3.72 & 11.1 & $6.09  \pm  3.28$  & D\\
61394  & SDF J132456.5+272212 & 5.761 & 39.6 & 14.4 & 7.09 & $3.45  \pm  0.47$  & F\\
62511  & SDF J132419.6+272228 & 5.665 & 15.0 & 5.23 & 4.81 & $6.52  \pm  2.07$  & D\\
65899  & SDF J132438.0+272315 & 5.687 & 15.0 & 5.29 & 5.34 & $13.8  \pm  2.13$  & D\\
66162  & SDF J132528.8+272317 & 5.688 & 16.4 & 5.78 & 8.01 & $12.0  \pm  2.56$  & D\\
70773  & SDF J132347.1+272414 & 5.696 & 6.77 & 2.39 & 11.0 & $3.35  \pm  2.86$  & D\\
71751  & SDF J132344.8+272427 & 5.720 & 10.4 & 3.70 & 9.96 & $5.30  \pm  1.27$  & F\\
75550  & SDF J132447.7+272512 & 5.708 & 14.9 & 5.28 & 6.63 & $4.74  \pm  3.44$  & D\\
81382  & SDF J132451.4+272626 & 5.697 & 9.05 & 3.20 & 8.55 & $6.20  \pm  2.00$  & D\\
83092  & SDF J132441.3+272649 & 5.747 & 15.6 & 5.62 & 5.50 & $5.70  \pm  5.22$  & F\\
89624  & SDF J132424.8+272812 & 5.714 & 13.1 & 4.67 & 8.36 & $4.08  \pm  5.53$  & F\\
91179  & SDF J132417.8+272833 & 5.712 & 13.3 & 4.74 & 14.0 & $8.31  \pm  1.35$  & D\\
98040  & SDF J132432.9+273009 & 5.674 & 25.9 & 9.06 & 11.0 & $7.62  \pm  1.80$  & F\\
98461  & SDF J132428.3+273012 & 5.697 & 11.3 & 3.99 & 3.99 & $3.65  \pm  1.81$  & F\\
120018 & SDF J132340.6+273447 & 5.684 & 38.8 & 13.6 & 11.3 & $10.0  \pm  2.00$  & D\\
124905 & SDF J132352.9+273549 & 5.694 & 6.88 & 2.43 & 5.07 & $6.93  \pm  4.93$  & D\\
138608 & SDF J132450.9+273839 & 5.669 & 19.7 & 6.88 & 11.4 & $4.56  \pm  10.3$  & F\\
138624 & SDF J132357.4+273836 & 5.677 & 8.30 & 2.60 & 11.0 & $5.04  \pm  2.75$  & D\\

\enddata
\tablenotetext{a}{ID in the NB816-detected catalog of S06}
\tablenotetext{b}{The redshift was derived from the wavelength of the flux peak in an observed spectrum assuming the rest wavelength of Ly$\alpha$ to be $1215$\AA.
These measurements could be overestimated in the case of significant damping wings by IGM.
Also, the observed peak position was slightly shifted redward due to instrumental resolution. See Figure \ref{fig_prof}.}
\tablenotetext{c}{The observed line flux corresponds to the total amount of the flux within the line profile. The slit-loss was corrected.}
\tablenotetext{d}{No dust absorption correction was applied.}
\tablenotetext{e}{Corrected for instrumental broadening.}
\tablenotetext{f}{Observed with FOCAS (F) or DEIMOS (D).}

\end{deluxetable*}

\section{Summary of the previous LAE Sample at $z=6.5$ and $z=5.7$ as of 2006}

The sample selection of photometric LAE candidates at $z=6.5$ and $z=5.7$ was presented in \citet{tan05} and S06, respectively.
The details of observation, photometry and color selection were presented in those papers; here, we briefly discuss about our photometric selection of the LAE sample.
The sample was based on flux-excess objects in narrow-band $NB921$ and $NB816$ images, compared with very deep broadband images of the SDF \citep{kas04}.
We selected LAE candidates at $z=6.5$ with definite NB excesses down to the $5\sigma$ limiting magnitude of $NB921=26.0$, {\it i.e.}, $z'-NB921>1.0$ and $z'-NB921>(z'-NB921)_{3\sigma}$, where ($z'-NB921$)$_{3\sigma}$ is the $3\sigma$ error in the color of $z'-NB921$, and the very red color in broad-bands, {\it i.e.}, ($i'-z' \geq1.3$ and $z<i'_{2\sigma}-1.3$) or ($z \geq i'_{2\sigma}-1.3$), where $i'_{2\sigma}$ are defined as $2\sigma$ limiting magnitudes of $i'$, to reduce contamination from foreground emission-line galaxies.
We also applied the no detection (ND; $ \leq 3\sigma$) criteria in all bands blueward of the dropout band, {\it i.e.}, $B$, $V$, and $R$.
In \citet{tan05}, we found $58$ photometric candidates at $z=6.5$ down to $NB921=26.0$ in the effective survey region of $876$ arcmin$^2$; spectroscopic confirmations of $17$ candidates were presented in K06.
In S06, we selected LAE candidates at $z=5.7$ with NB excesses with $i'-NB816\geq1.5$ down to $NB816=26.0$ and very red color in broad-bands, {\it i.e.}, ($R-z' \geq1.5$ and $R<R_{2\sigma}$) or ($R \geq R_{2\sigma}$), where $R_{2\sigma}$ are defined as $2\sigma$ limiting magnitudes of $R$.
We also applied the ND ($ \leq 2\sigma$) criterion in the $B$ and $V$ bands.
The $89$ photometric candidates were found at $z=5.7$ down to $NB816=26.0$ in the same survey region; $34$ of them were confirmed by spectroscopy.
We estimated the detection completeness, defined as the ratio of detected LAE candidates to all the LAEs actually present in the universe.
This should be corrected in evaluating the LF as a function of $NB$ magnitudes by counting artificial objects distributed on the real $NB$ images.
The detection completeness was evaluated as $>0.8$ at $NB816<25.0$ and $\sim0.75$ at $NB816=26.0$, and as $>0.75$ at $NB921<25.0$ and $\sim0.45$ at $NB921=26.0$, for the $z=5.7$ and $6.5$ sample, respectively.
We quantitatively distinguished LAEs from nearby emitters based on their asymmetric emission-line profile, using the ^^ ^^ weighted skewness" indicator (K06).
The comoving survey volume was as large as $2.17\times10^5$ Mpc$^3$ and $1.80\times10^5$ for $z=6.5$ and $z=5.7$, respectively.
It should be noted that the LAE samples at these two epochs were extracted from the same field using the same photometric procedure, similar survey volume, and similar selection criteria.
We carefully determined the $NB$-excess criteria to provide almost the same EW thresholds (EW$_0>10$\AA) for both LAE samples.

\section{New spectroscopic identifications from LAE sample}

\subsection{Spectroscopic Observations}

We carried out further spectroscopic observations with the Subaru FOCAS \citep{kas02} and the Keck II DEIMOS \citep{fab03} over the last four years (2006---2010).
This spectroscopic campaign is summarized in Table~\ref{tab_obs}.

The FOCAS observations were made with a $300$-line mm$^{-1}$ grating and an O58 order-cut filter.
The spectra covered $5400-10,000$ \AA, with a pixel resolution of $1.34$ \AA.
The $0\arcsec.6$-wide slit gave a spectroscopic resolution of $7.1$ \AA~($R\sim1300$).
The spatial resolution was $0\arcsec.3$ pixel$^{-1}$ with $3$-pixel on-chip binning.
We also allocated slits for $NB921$-strong ($z'-NB921>1$) emitters, irrespective of their ($i'-z'$) color, as a LAE criterion to test our selection criteria.
Some extra slits were allocated for our candidates in another observation by \citet{nag07}, in which they used a $175$-line mm$^{-1}$ Echelle grating, an SDSS $z'$ filter as an order-cut filter, and $0\arcsec.83$ slit, giving $R\sim1500$.
Extra slits were also allocated for observations by M.Iye et al. (in preparation), in which they used the VPH900 grism, O58 order-cut filter, and $0\arcsec.8$ slit, giving $R\sim1500$.
We obtained spectra of standard stars Hz 44, Feige 34, and BD+28 for flux calibration.
The data were reduced in a standard manner.

For the DEIMOS observations, we used an $830$-line mm$^{-1}$ grating and a GG495 order-cut filter.
The central wavelength was set to $\lambda_c=8100$ \AA.
The slit width was $1\arcsec.0$ with $0.47$ \AA~pixel$^{-1}$, giving a resolving power of $\sim3600$.
The wavelength coverage was $\sim5000---10,000$ \AA, depending on the position in the mask.
We also allocated slits for strong $NB921$ emitters.
Some extra slits were allocated for our candidates in another observation led by L. Jiang et al. (2011, in preparation), in which they used almost the same instrumental setup, except that $\lambda_c=9239$ \AA.
We obtained spectra of standard stars BD +28 4211 and Hz 44 for flux calibration.
The data were reduced with the spec2d pipeline\footnote{The data reduction pipeline was developed at University of California, Berkeley, with support from National Science Foundation grant AST 00-71048.} for DEEP2 DEIMOS data reduction.
During the 2010 run, the flexure compensation system (FCS) of DEIMOS was broken.
Unexpected flexure at this time was adequately corrected by shifting the spectral images based on the instrumental rotation angle, which is sensitive to the flexure change, of each exposure.

All spectroscopic observations were taken at high enough resolving power to distinguish single Ly$\alpha$ emission from $[$O {\sc ii}$]$ doublets (rest-frame separation of $2.78$\AA); however, in practice, it is hard to discriminate between the features, given very faint emissions.
To quantitatively distinguish LAEs from nearby emitters, we used a line asymmetry estimator, called weighted skewness, $S_w$, as in K06.
Skewness is a popular statistic defined as the third moment of the distribution function.
In K06, we found that this estimator is sensitive to asymmetry and carefully determined the critical value to distinguish LAEs from nearby emitters based on our $NB921$- and $NB816$-excess sample.
The $S_w$ values of foreground emitters never exceeded $S_w=3$, which we set as the critical $S_w$ value to distinguish LAEs from foreground emitters.
We should note that the threshold $S_w=3$ applied in this study is only valid for LAEs at $z=5.7$ and $6.5$.
We have no guarantee that the same critical value can be used for LAEs at different $z$, where, in principle, different intergalactic attenuation could change the degree of asymmetry in the Ly$\alpha$ emission.
We serendipitously identified some LAEs (three for $z=6.5$ and eight for $z=5.7$) from strong NB921 and NB816 emitters that did not meet our color selection criteria.
We found that most of them had very close neighbors in the images, which prevented accurate aperture photometry.
In summary, we identified $28$ and $20$ additional LAEs at $z=6.5$ and $5.7$, respectively\footnote{SDF J132417.9+271746 and SDF J132521.1+272712, classified as single-line emitters in \citet{tan05}, and SDF J132417.8+272833 (ID 91179), classified as a single-line emitter in S06, were found to be LAEs in the deep spectroscopic observations performed in this study.}.
The spectra of these additional LAEs are shown in Figures \ref{fig_spe65} and \ref{fig_spe57}, and their spectroscopic properties are summarized in Tables \ref{tab_laespec65} and \ref{tab_laespec57} for $z=6.5$ and $5.7$, respectively.

\begin{figure}
\vspace*{-1.3cm}
\hspace*{-1.0cm}
\plotone{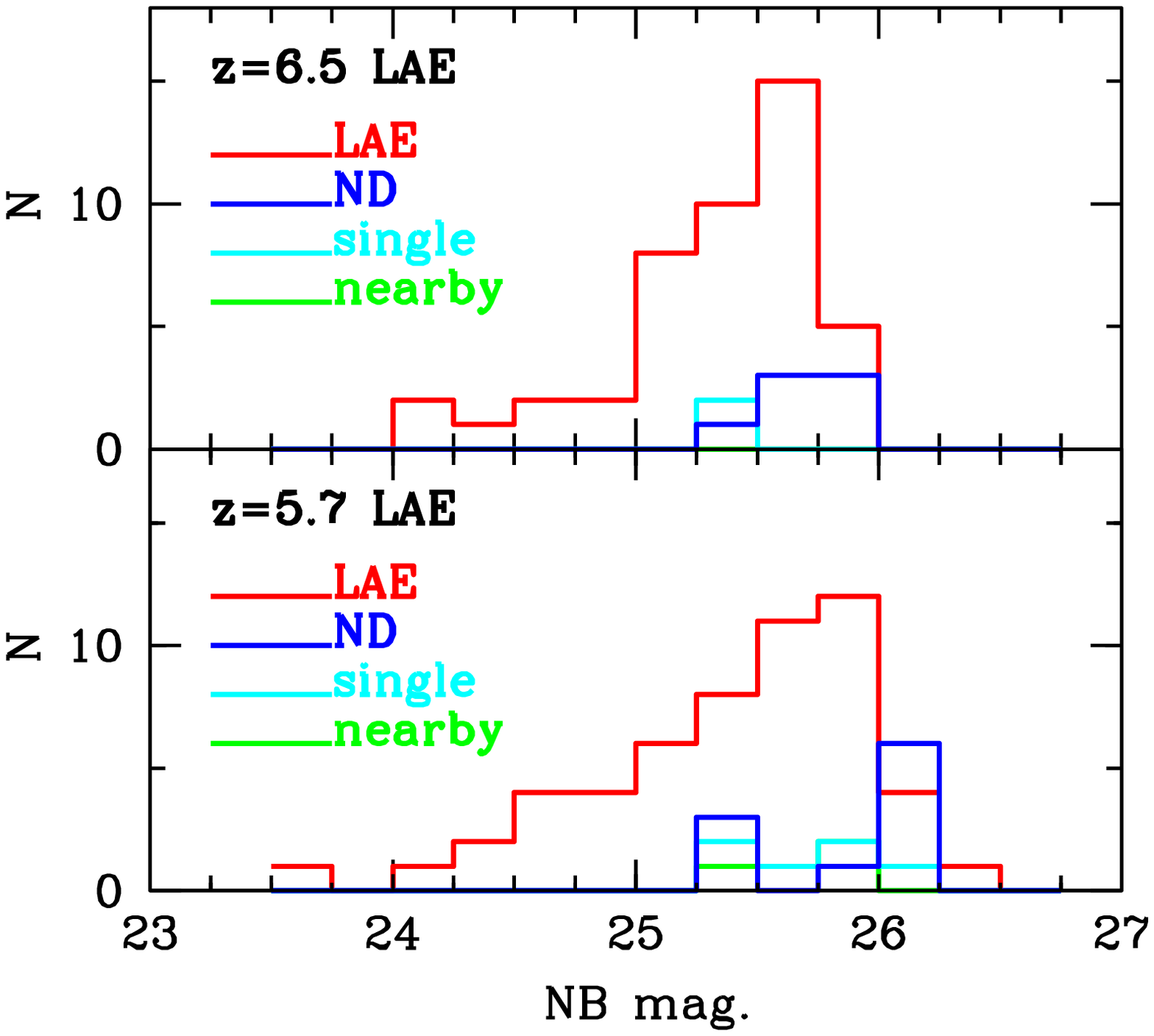}
\epsscale{1.0}
\vspace*{-0.5cm}
\caption{
Magnitude histogram of spectroscopically identified LAEs.
\label{fig_maghist}}
\end{figure}

We did not detect N$_{\sc V}$ $\lambda1240$, which is the only accessible strong high-ionization metal line indicative of AGN activity, from any of these newly identified LAEs.
We have some candidates (two for $z=6.5$ and six for $z=5.7$) that exhibited only a faint, poorly fit ^^ ^^ single" line emission signal, so we could not verify its asymmetry, which is the only key diagnostic for distinguishing Ly$\alpha$ emission from other emissions. 
We did not include these single-line emitters in the LAE samples.
We had some additional candidates ($7$ for $z=6.5$ and $10$ for $z=5.7$) for which no emission signal was detected, even using $8-10$m class telescopes. 
These might be intrinsically too faint, or some could be transient objects.
Table~\ref{tab_sum} summarizes our current spectroscopic identifications of LAEs at $z=6.5$ and $5.7$ in the SDF and Figure~\ref{fig_maghist} shows the NB-magnitude histogram of spectroscopic identifications.
At this time, $90\%$ and $74\%$ of the photometric candidates at $z=6.5$ and $5.7$, respectively, have been followed by spectroscopy.
Uncertain ^^ ^^ single" and ^^ ^^ no-detection (ND)" candidates are of course only dominant at the very faint end of the sample, so we consider this to be our current observational feasibility limit.
Despite this uncertainty, most of the photometric candidates have been spectroscopically identified, and only six candidates remain without follow-up spectroscopy in the $z=6.5$ sample.

\epsscale{1.5}

\subsection{Spectroscopic Sample}

\begin{figure}
\vspace*{-1.3cm}
\hspace*{-0.6cm}
\plotone{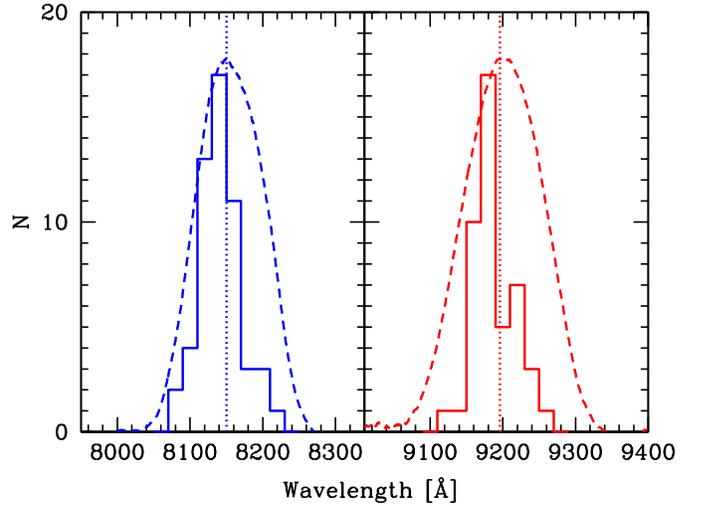}
\epsscale{1.0}
\vspace*{-0.3cm}
\caption{
Observed wavelength distribution of the Ly$\alpha$ line peak for our spectroscopic sample at $z=5.7$ (left) and $6.5$(right).
The transmission curves of $NB816$ (left) $NB921$ (right) are overplotted as dashed lines.
\label{fig_peak}}
\end{figure}

Now the total numbers of spectroscopically confirmed LAEs in the SDF are $45$ and $54$ at $z=6.5$ and $5.7$, respectively.
Figure~\ref{fig_peak} shows the line-peak wavelength distribution of spectroscopically confirmed LAEs.
The distributions of line peaks show an apparent blueward deviation against the NB transmission curve, which was already confirmed in our previous studies.
This is due to the fact that the asymmetric LAE profile, with a broad red wing and sharp blue cutoff and a discontinuous Lyman continuum break, contributes more to the NB flux when a line peak lies at the blue side of the transmission curve.
The evaluations of sample completeness and contamination rates are listed in Table~\ref{tab_sum}.
At least $81\%$ ($70\%$) of our photometric candidates at $z=6.5$ ($5.7$) have been spectroscopically identified as real LAEs.
Because ^^ ^^ single" and ^^ ^^ ND" candidates depend on uncertain assumptions, we evaluated two extreme values, {\it i.e.}, we assumed that all of these uncertain candidates were really LAEs, and then assumed that none of the candidates was really a LAE.
By taking the average of these two cases, our photometric sample completeness factor, determined as the ratio of the number of true LAEs to the number of objects that meet our selection criteria ($=(1-{\rm \langle CT \rangle})/{\rm \langle CP \rangle}$, see Table~\ref{tab_sum}), was evaluated to be $(1-0.105)/0.935=0.957$ and $(1-0.180)/0.860=0.953$ for $z=6.5$ and $z=5.7$, respectively.
These spectroscopic results indicate that our photometric sample has high reliability, indicating our color selection process works well.
$[$O {\sc iii}$]$ and $[$O {\sc ii}$]$ emitters identified by spectroscopy based on their doublet signatures, were removed from the photometric sample, whereas LAEs that were serendipitously found by spectroscopy but not listed in the original photometric sample were included in the following analysis.

\epsscale{1.25}

\begin{deluxetable*}{lllllllllllllll}
\tabletypesize{\scriptsize}
\tablecaption{Summary of Spectroscopic Identifications\label{tab_sum}}
\tablewidth{0pt}
\tablehead{
\colhead{z} & \colhead{N$_{\rm phot}$\tablenotemark{a}} & \colhead{Ly$\alpha$} & \colhead{H$\alpha$} & \colhead{$[$O {\sc iii}$]$} & \colhead{$[$O {\sc ii}$]$} & \colhead{single\tablenotemark{b}} & \colhead{ND\tablenotemark{c}} & \colhead{wo/spec\tablenotemark{d}} & \colhead{Ly$\alpha_{\rm serend}$\tablenotemark{e}} & \colhead{Ly$\alpha_{\rm total}$\tablenotemark{f}} & \colhead{CP$_{\rm min}$\tablenotemark{g}} & \colhead{CP$_{\rm max}$\tablenotemark{h}} & \colhead{CT$_{\rm max}$\tablenotemark{i}} & \colhead{CT$_{\rm min}$\tablenotemark{j}}
}
\startdata
6.5 & 58 & 42 & 0 & 1 & 0 & 2 & 7  & 6  & 3 & 45 & 93\% & 94\% & 19\% & 2\% \\
5.7 & 89 & 46 & 0 & 0 & 4 & 6 & 10 & 23 & 8 & 54 & 85\% & 87\% & 30\% & 6\% \\
\enddata
\tablenotetext{a}{Photometric candidates}
\tablenotetext{b}{These possess only a single poor-quality symmetric emission feature with small $S_w<3$, in which we cannot determine whether they are Ly$\alpha$ emission or $[$O {\sc ii}$]$ emission (K06).}
\tablenotetext{c}{No detection of any emission features}
\tablenotetext{d}{Candidates without spectroscopic followup observations}
\tablenotetext{e}{LAEs serendipitously discovered in spectroscopic observation. They are not included in the photometric candidates.}
\tablenotetext{f}{Total number of LAEs spectroscopically identified $=$Ly$\alpha+$Ly$\alpha_{\rm serend}$}
\tablenotetext{g}{Sample completeness rate provided that all singles are foreground objects $=$Ly$\alpha/$Ly$\alpha_{\rm total}$}
\tablenotetext{h}{Sample completeness rate provided that all singles are LAEs $=$(Ly$\alpha+$single)/(Ly$\alpha_{\rm total}+$single)}
\tablenotetext{i}{Sample contamination rate provided that all singles and NDs are foreground objects $=$($[$O {\sc iii}$]$,$[$O {\sc ii}$]$+single+ND)/(N$_{\rm phot}$-wo/spec)}
\tablenotetext{j}{Sample contamination rate provided that all singles and NDs are LAEs $=[$O {\sc iii}$]$,$[$O {\sc ii}$]$/(N$_{\rm phot}$-wo/spec)}
\end{deluxetable*}

\section{Ly$\alpha$ and rest-UV luminosity functions}

\subsection{Deriving Ly$\alpha$ Luminosity}

We derived Ly$\alpha$ luminosity both photometrically and spectroscopically.
The latter would be more accurate than the former, but hardly gives an accurate measurement of faint continuum flux in most cases, though these are eventually found to be fairly consistent with one another.
We measured luminosities more carefully in this than the previous study (K06) to accurately determine Ly$\alpha$ and rest-UV LFs.
We applied the following procedure, not only to the newly identified LAE sample but also to the previously identified objects by recalculating the Ly$\alpha$ and rest-UV luminosities.

First, for the photometric estimate of Ly$\alpha$ line flux, the Ly$\alpha$ line ($f_{\rm line}$; erg s$^{-1}$ cm$^{-2}$) and rest-UV continuum ($f_c$; erg s$^{-1}$ cm$^{-2}$ Hz$^{-1}$) fluxes were evaluated from NB and BB magnitudes ($m_{NB}$ and $m_{BB}$) as follows:

\begin{eqnarray}
m_{NB, BB}+48.6 =-2.5{\rm log} \frac{\int^{\nu_{Ly\alpha}}_{0}(f_c+f_{\rm line})T_{NB, BB} d\nu/\nu}{\int T_{NB, BB} d\nu/\nu}, 
\end{eqnarray}

\noindent where $\nu_{Ly\alpha}$ is the observed frequency of Ly$\alpha$, and $T_{NB}$ and $T_{BB}$ are the transmission bandpasses of the NB and BB filters, respectively.
The (NB, BB) filter combinations in the above formula are ($NB921$, $z'$) and ($NB816$, $z'$) for $z=6.5$ and $5.7$, respectively.
We assumed that the spectral energy distribution (SED) of LAEs had a constant $f_c$ and $\delta$-function Ly$\alpha$ emission profile, whose line-width was much smaller than $\nu_{Ly\alpha}$.
The estimate of the Ly$\alpha$ line flux was not largely affected, even when taking account of a finite line width as of observed $\lesssim30$\AA, which made $\sim16\% $ Ly$\alpha$ flux change at most at the reddest wavelength of the NB-band.
The observed value was adopted for $\nu_{Ly\alpha}$ when spectroscopically identified; otherwise, the central frequency, $\nu_c$, of the NB filters was used.
We used the actual filter transmission curves convolved by telescope, instrument and atmospheric transmissions for $T_{NB, BB}$.
The importance of such a careful analysis of filters' non-square transmission curve has been pointed out in \citet{gro07}.
If an object was not detected in the BB filter, $m_{BB}$ was replaced by the $1\sigma$ limiting magnitude of BB, and $f_c$ was forcibly set to $0$ when $f_c<0$.

In spectroscopic measurements of Ly$\alpha$ line flux, we corrected for slit loss.
The target was regarded as almost perfectly centered in the slit because the typical positioning error of the slit on the target object was as small as $0\arcsec.1$ rms for both FOCAS and DEIMOS spectrographs.
Thus, we assumed that the slit flux loss depended only on the seeing size, target size, and slit width. 
We calculated a possible slit flux loss based on the seeing size and slit width for each spectroscopic observation.
An intrinsic Ly$\alpha$ radial light profile of each object was simply assumed to be a Gaussian profile with FWHM measured on the NB image deconvolved by the seeing size ($0\arcsec.98$) of the image.
The slit lengths were always so much longer ($>8\arcsec.0$) than the object size that slit loss along the spatial direction was negligible.
The resultant silt loss was corrected for the spectroscopically measured Ly$\alpha$ flux for each object.
The average slit loss was evaluated to be $24.1\pm10.1\%$ and $30.3\pm12.5\%$ for $z=6.5$ and $5.7$, respectively.
The difference in slit flux loss between these two samples was minor.

\begin{figure}
\hspace*{-1.0cm}
\plotone{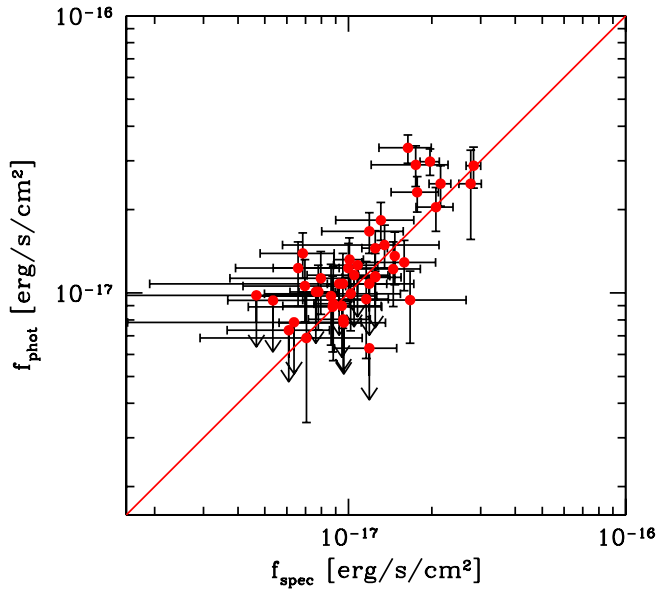}
\vspace*{-1.5cm}
\epsscale{1.00}
\caption{
Comparison of the Ly$\alpha$ fluxes measured by spectra, $f_{spec}$, with those inferred from photometry, $f_{phot}$, for our spectroscopic LAE sample at $z=6.5$.
The solid line represents a one-to-one correspondence between f$_{spec}$ and f$_{phot}$.
The errors were estimated based on the sky rms fluctuation of each spectrum for f$_{spec}$ and errors in magnitudes for f$_{phot}$.
\label{fig_flaecomp_65}}
\end{figure}

\begin{figure}
\hspace*{-1.0cm}
\plotone{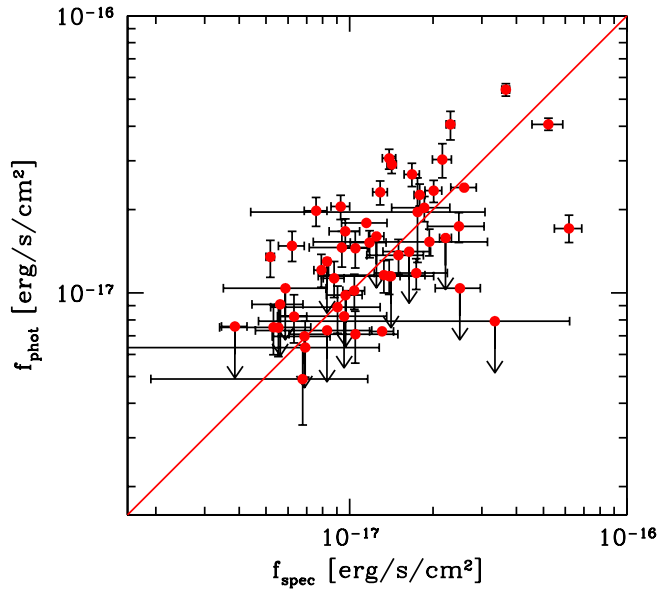}
\vspace*{-1.5cm}
\epsscale{1.00}
\caption{
Same as Figure~\ref{fig_flaecomp_65}, but for LAEs at $z=5.7$.
\label{fig_flaecomp_57}}
\end{figure}

Comparisons of the Ly$\alpha$ fluxes measured photometrically and spectroscopically for the spectroscopically identified LAE sample are shown in Figures~\ref{fig_flaecomp_65} and~\ref{fig_flaecomp_57}.
The correspondence is good except in a few cases.
Note that the photometric Ly$\alpha$ flux may have non-negligible uncertainty when $m_{BB}$ is only given by an upper limit, as appears often in the $z=6.5$ sample.
We hereafter use spectroscopically measured Ly$\alpha$ fluxes for the spectroscopically identified LAEs, and photometrically inferred Ly$\alpha$ fluxes for the remaining uncertain (referred to as ^^ ^^ single", ^^ ^^ ND", and ^^ ^^ wo/spec" in Table~\ref{tab_sum}) objects.

\epsscale{1.35}

\epsscale{1.35}

\subsection{Ly$\alpha$ Luminosity Function}

We estimated the acceptable Ly$\alpha$ LF range specified by the upper and lower limits.
The upper limit was estimated assuming that all the uncertain (^^ ^^ single", ^^ ^^ ND", and ^^ ^^ wo/spec" in Table~\ref{tab_sum}) photometric candidates, for which photometrically evaluated Ly$\alpha$ luminosities were adopted, were really LAEs.
The lower limit was estimated assuming that all the uncertain candidates were not LAEs, {\it i.e.}, using only the pure spectroscopically identified LAE sample.
To derive the Ly$\alpha$ LF, the detection completeness, which decreases at fainter NB magnitudes, was corrected in the same way as outlined in K06.
In both the upper and lower limit estimates, we corrected for this detection completeness by number weighting according to the NB magnitude.

\begin{figure}
\vspace*{-2.2cm}
\hspace*{-0.6cm}
\plotone{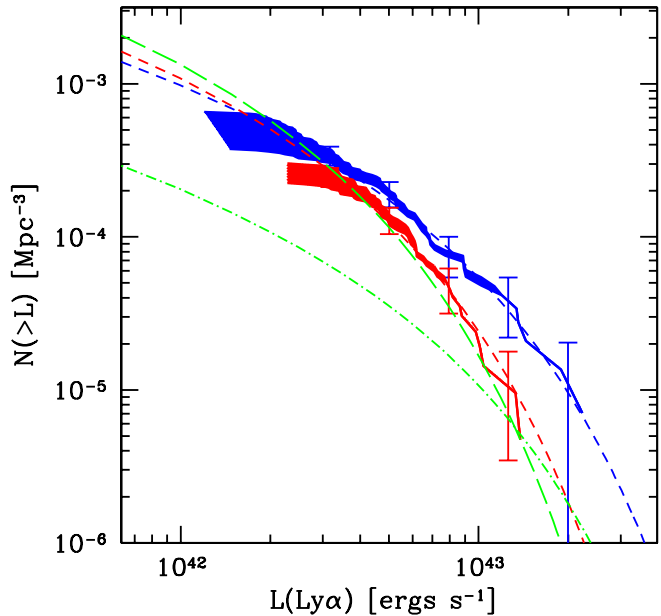}
\epsscale{1.00}
\caption{
Comparison of the cumulative Ly$\alpha$ LFs of LAEs at $z=6.5$ (red-shaded region) and $z=5.7$ (blue-shaded region).
We estimated the acceptable Ly$\alpha$ LF ranges as specified by the upper and lower limits.
The upper limit was estimated assuming that all the uncertain (referred to as ^^ ^^ single", ^^ ^^ ND", and ^^ ^^ wo/spec" in Table~\ref{tab_sum}) photometric candidates are really LAEs, and the lower limit was estimated assuming that all the uncertain candidates are not LAEs, {\it i.e.}, using only the pure spectroscopically identified LAE sample.
In both the upper and lower limit estimates, we corrected for the detection completeness by number weighting according to the NB magnitude.
Error bars evaluated by the Poisson errors are shown in some average data points between the upper limits and lower limits.
The short-dashed lines (red for $z=6.5$ and blue for $z=5.7$) show the fitted Schechter LFs in the case of $\alpha=-1.5$.
As a comparison, the green long-dashed line shows the Ly$\alpha$ LF at $z=6.5$ from \citet{ouc10}, and the green dot-dashed line shows that of \citet{hu10}.
\label{fig_lalf}}
\end{figure}

Figure~\ref{fig_lalf} provides a comparison of the cumulative Ly$\alpha$ LFs between $z=6.5$ and $5.7$.
The red-shaded and blue-shaded regions indicate acceptable LF ranges for $z=6.5$ and $5.7$, respectively.
Compared with our previous estimate in K06, the acceptable LF ranges are now more sharply determined.
The given uncertainties are by the Poisson errors, shown for some average data points between upper limit and lower limit data points, indicating that Poisson errors are dominant at the bright end, whereas the spectroscopic uncertainties are dominant at the faint end.
Taking into account the corrections for completeness factor ($95.7\%$ for $z=6.5$ and $95.3\%$ for $z=5.7$) evaluated in the previous section, both LFs could increase by a factor of $\sim1.05$, although this is smaller than the Poisson errors.
Note that we derived the apparent Ly$\alpha$ luminosity uncorrected for either dust extinction or self-absorption, which is evident on the blue-side cutoff of the emission line.

\epsscale{1.45}

We fitted a Schechter function, $\phi(L)dL=\phi^*(L/L^*)^\alpha$ exp$(-L/L^*)dL/L^*$, to the average data points between the upper limit and lower limit data points.
The $\chi^2$ was minimized with a single grid search in the two parameters, $L^*$ and $\phi^*$, for fixed slopes of $\alpha=-1.7$, $-1.5$, and $-1.3$.
The larger value of the Poisson errors or the difference between the upper limit and the lower limit was used for the error of each data point when fitting a Schechter function.
The derived best-fit parameters are listed in Table~\ref{tab_sch}.

\begin{deluxetable}{llll}
\tabletypesize{\footnotesize}
\tablecaption{Best-fit Schechter Parameters for LAE LF at $z=6.5$ and $5.7$ of the SDF\label{tab_sch}}
\tablewidth{0pt}
\tablehead{
\colhead{Sample} & \colhead{$\alpha$} & \colhead{$L^*$}   & \colhead{$\phi^*$} \\
                 &         (fix)      & log(/$h_{70}^{-2}$ ergs s$^{-1}$) & log(/$h_{70}^{3}$ Mpc$^{-3}$)
}
\startdata
z = 6.5 & -1.7 & $42.82^{+0.10}_{-0.10}$ & $-3.40^{+0.22}_{-0.18}$ \\
        & -1.5 & $42.76^{+0.10}_{-0.10}$ & $-3.28^{+0.20}_{-0.20}$ \\
        & -1.3 & $42.70^{+0.10}_{-0.10}$ & $-3.20^{+0.20}_{-0.18}$ \\
\tableline
z = 5.7 & -1.7 & $43.12^{+0.06}_{-0.04}$ & $-3.74^{+0.06}_{-0.08}$ \\
        & -1.5 & $43.02^{+0.06}_{-0.06}$ & $-3.56^{+0.08}_{-0.10}$ \\
        & -1.3 & $42.94^{+0.06}_{-0.06}$ & $-3.34^{+0.10}_{-0.08}$ 
\enddata
\end{deluxetable}

After spectroscopic confirmations of a large amount of our LAE samples at both $z=6.5$ and $5.7$, we found that the evaluation of the Ly$\alpha$ LF at $z=6.5$ was apparently deficient compared with that of $z=5.7$, at least at the bright end, where the entire samples at both epochs have been spectroscopically identified.
A possible decline in the $z=6.5$ LF appeared even at the faint end.
The $L^*$ difference between $z=6.5$ and $z=5.7$ decreased slightly to $\sim 0.65$ magnitudes from the previous estimate of $\sim0.75$ magnitudes, given $\alpha=-1.5$.
A major cause of this change could be the significant increase in the number of spectroscopically confirmed LAEs and the more careful measurements of luminosity performed in the present study.

\subsubsection{Uncertainties in the LF estimates}

Two uncertainties appeared in the photometric estimate of the Ly$\alpha$ luminosity: the photometric error and the redshift ambiguity.
In the previous section, we assumed that the peak of the Ly$\alpha$ emission line was the central wavelength, $\lambda_c$, of the NB filters for the photometric sample.
We carried out a Monte Carlo simulation to see how the associated uncertainties would affect the resultant LF.
In the simulations, Gaussian random photometric errors both in BB and NB were assigned to the measured magnitudes of the photometric sample, and redshifts were assigned so that their redshift distributions matched the observations, as shown in Figure~\ref{fig_peak}.
The $1\sigma$ photometric error was estimated from the background fluctuation and the flux Poisson noise.
The process of recalculating the LF and deriving the best-fit Schechter parameters was repeated many times.
We found rms fluctuations of $\sigma( {\rm log}(L^*))=0.041$ and $\sigma( {\rm log}(\phi^*))=0.066$ for $z=6.5$ and $\sigma( {\rm log}(L^*))=0.023$ and $\sigma( {\rm log}(\phi^*))=0.027$ for $z=5.7$ for $\alpha=-1.5$, suggesting that these uncertainties have only a small impact on the result.
The photometric sample selection prior to the luminosity measurements might be affected by errors in the photometric catalog.
We also carried out a Monte Carlo simulation to see how the photometric error in the catalog would affect the color selection and resultant LF, by assigning random errors to the measured magnitudes of detected object catalogs in all bands.
We did not use spectroscopic estimate of the Ly$\alpha$ flux in the simulation, and estimated the Ly$\alpha$ LFs photometrically.
The rms fluctuations were found to be $\sigma( {\rm log}(L^*))=0.104$ and $\sigma( {\rm log}(\phi^*))=0.119$ for $z=6.5$ and $\sigma( {\rm log}(L^*))=0.098$ and $\sigma( {\rm log}(\phi^*))=0.094$ for $z=5.7$ for $\alpha=-1.5$.
We note that more than 70\% of our LAE samples have been spectroscopically identified, whereas we have to add the redshift ambiguity in all the sample in this simulation; therefore, the uncertainties of the LF estimates are apparently overestimated.

Given an observed NB flux, the photometric Ly$\alpha$ flux tended to be overestimated, and the continuum flux tended to be underestimated when the actual line peak wavelength was smaller than the $\lambda_c$ due to a strong Lyman break in the continuum, as well as to the asymmetric Ly$\alpha$ line profile.
This trend would be expected if the transmission curve of the NB filter was nearly a top-hat shape; however, we used NB filters with almost Gaussian-shaped transmission curves, as shown in Figure~\ref{fig_peak}.
For these filters, photometrically estimated line flux would maintain the observed NB excess by decreasing when the line peak shifted farther away, whether it was redder or bluer, from the $\lambda_c$. 
As a result, the photometric line flux was largely underestimated when an emission line was really located in the redder part of the NB transmission, whereas in the blue portion, an underestimate due to the low transmission was balanced by an overestimate due to the continuum Lyman break.
We confirmed these trends by numerical experiment.
As the observed line-peak distribution deviated to blue, the systematic error in the photometric estimate of Ly$\alpha$ line would be small.
In our Monte Carlo simulation, we found that the systematic error caused by this trend was as small as $\sigma( {\rm log}(L^*))=0.04$ and $0.06$ for $z=5.7$, and $6.5$, respectively.
We also performed a Monte Carlo simulation to investigate any possible distortion that the discrepancy, as seen in Figure~\ref{fig_flaecomp_65} and \ref{fig_flaecomp_57}, between spectroscopically measured and photometrically inferred Ly$\alpha$ luminosities could have on the result. 
When Gaussian random error with the same scatter as in Figure~\ref{fig_flaecomp_65} and \ref{fig_flaecomp_57} was assigned to each Ly$\alpha$ luminosity, the best-fit Schechter parameters, given $\alpha=-1.5$, only changed by $\sigma( {\rm log}(L^*))=0.023$ and $\sigma( {\rm log}(\phi^*))=0.042$ for $z=6.5$ and $\sigma( {\rm log}(L^*))=0.042$ and $\sigma( {\rm log}(\phi^*))=0.070$ for $z=5.7$, which were negligible.

\subsubsection{Comparisons with other studies}

The Schechter parameters derived in this study are almost identical to previous estimates by S06 for $z=5.7$, whereas at $z=6.5$, the values in this study fell between the previous estimates of the upper and lower limits.
These parameters are also consistent, within the errors, with independent studies by \citet{ouc10} and \citet{ouc08} for $z=6.5$ and $5.7$.
The estimates at $z=6.5$ given by \citet{hu10} differed by the largest amount from our estimate, especially at faint luminosities.
Their estimated Ly$\alpha$ LF was almost a factor of three (five) lower than estimates of K06, \citet{ouc08} and \citet{ouc10} at their faintest bin at log$L$(Ly$\alpha$)$\sim42.8$ at $z=6.5$ ($5.7$).
\citet{hu10} claimed that the difference might be mainly caused by a large contamination in the photometric sample, if present; however, the present study, based on a large number of spectroscopic confirmations, completely disallows this interpretation.
As shown in Section 3, the contamination of our sample was as low as $\leq20\%$, evaluated by the spectroscopic results, and the sample completeness was as high as $\geq85\%$ for both the $z=6.5$ and $5.7$ samples.
The photometric LAE selection of \citet{ouc08} and \citet{ouc10} nearly matches this study, so their sample is probably also less affected by contamination.
The reason for the difference in the two evaluations at the faint end is unclear because the data reduction, photometry, and the LAE selection differ slightly.
However, we can suggest one plausible explanation.
The Ly$\alpha$ LF at $z=6.5$ by \citet{hu10}, which is basically based on their spectroscopic sample, is very close to that of our spectroscopic estimate (K06).
Spectroscopic confirmation at the faint end is generally so difficult that our previous spectroscopic sample apparently lacked completeness at the faint end. 
The present study, based on spectroscopic observations with deeper typical integration times (10 ksec, see Table~\ref{tab_obs}) than \citet{hu10}, overcame the problem and revealed that most of these faint unidentified objects were real LAEs.
As almost half of the faint photometric sample of \citet{hu10} were not spectroscopically identified at $z=5.7$, the completeness difference at the faint end could explain the significant difference in resultant Ly$\alpha$ LFs.
Nonetheless, it is interesting to note that both studies, more or less, detected a difference in the Ly$\alpha$ LFs between $z=6.5$ and $5.7$

\epsscale{1.35}
\begin{figure}
\vspace*{-1.2cm}
\hspace*{-0.2cm}
\plotone{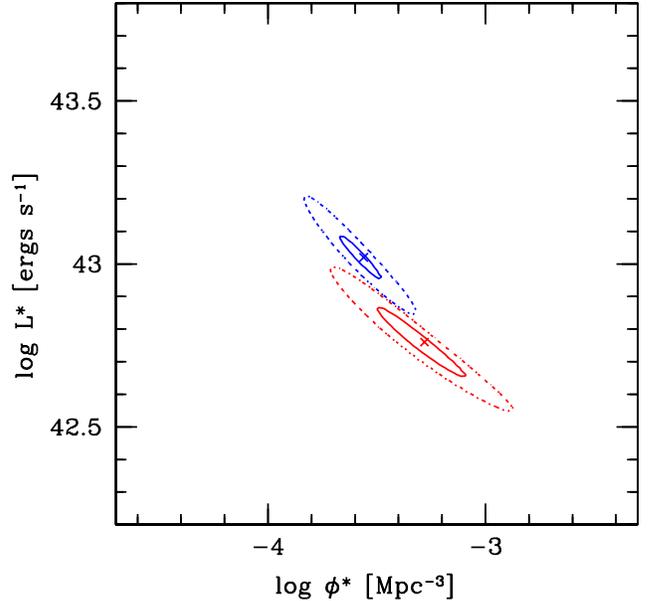}
\vspace*{-0.5cm}
\epsscale{1.00}
\caption{
Error ellipses of the best-fit Schechter parameters $\phi^*$ and $L^*$ of Ly$\alpha$ LF given a fixed $\alpha=-1.5$.
The lower ellipse (red) is for LAEs at $z=6.5$, and the upper ellipse (blue) is for $z=5.7$.
The inner and outer solid ellipses are the $1$ $\sigma$ and $3\sigma$ confidence levels, respectively.
\label{fig_eelip}}
\end{figure}

\subsubsection{Significance of the LF difference and the cosmic variance}

To illustrate the significance of the LF difference between $z=6.5$ and $5.7$, we plotted the error contours for our Schechter-parameter fits in Figure~\ref{fig_eelip}.
The confidence levels of the fitting were computed based on the larger of the Poissonian error or the difference between the upper limit and the lower limit.
Figure~\ref{fig_eelip} reveals that the ($L^*$, $\phi^*$) error ellipses at fixed $\alpha=-1.5$ for $z=5.7$ and $6.5$ do not overlap each other; that is, the difference in LF between $z=5.7$ and $6.5$ is significant at almost the $3\sigma$ level.
This is also the case for any $\alpha$.
The difference in $L^*$ is more significant than that in $\phi^*$.
Based on \citet{som04}, we evaluated the cosmic variance of our LAE samples. 
We assumed a one-to-one correspondence between LAEs and dark haloes, and used their predictions at $z=6$.
With our comoving survey volume of $2.17\times10^5$ $h_{70}^{-3}$ Mpc$^3$ and a number density of $2.76\times10^{-4}$ ($1.94\times10^{-4}$) $h_{70}^3$ Mpc$^{-3}$ for the upper (lower) limit estimate, we obtained a cosmic variance of $\sim32\%$.
We also estimated a variance of $\sim 20\%$ for the $z=5.7$ LAE sample.
As shown by the error bars in Figure~\ref{fig_eelip}, the $3\sigma$ error circles for the two epochs overlap each other when cosmic variance is included; however, our upper limit estimate still differed from the $z=5.7$ result at the $2\sigma$ level.
Although the difference in LF was large at the bright end and small at the faint end, we forcibly attempted to fit the LF at $z=6.5$ for the Schechter function at $z=5.7$ with either fixed $\phi^*$ or $L^*$, given $\alpha=-1.5$.
We obtained log $L^*=42.90^{+0.02}_{-0.14}$ given a fixed $\phi^*$, corresponding to a $24\%$ decrease in the Ly$\alpha$ luminosity, $L^*(z=6.5)=0.76L^*(z=5.7)$, and we obtained ${\rm log}(\phi^*)=-3.74^{+0.06}_{-0.24}$ given a fixed $L^*$, corresponding to a $34\%$ decrease in LF amplitude, $\phi^*(z=6.5)=0.66\phi^*(z=5.7)$.
These values are comparable to the cosmic variance, indicating that the observed LF difference might be caused by the cosmic variance.
We cannot rule out this possibility, but we discuss it further in Section 8.



\subsection{Rest-UV Continuum Luminosity Function}

We derived the rest-UV continuum LFs of our LAE sample at $z=6.5$ and $5.7$.
In Section 4.1, the flux of the rest-UV continuum ($f_c$) was simultaneously derived once the Ly$\alpha$ line flux ($f_{\rm line}$) was determined by either spectroscopy or photometry.
The effective wavelength of the derived UV luminosity was $1250-1270$\AA~ for both $z=5.7$ and $6.5$.
The detection completeness in the $z'$ band, which corresponds to the rest-UV flux, should be corrected when calculating the UV LF; however, our LAE samples were basically selected in NB magnitude.
It is impossible to evaluate the detection completeness in the $z'$ band for the NB-selected sample; therefore, we should note that it is inevitable that the derived rest-UV continuum LF may be affected, especially at the faint end of the LF, by the difference in completeness of the NB and $z'$ bands.
As in our previous study, the correction was made based on the detection completeness in the NB filter evaluated in the previous section.
To overcome the problem, we take an alternative approach to derive the rest-UV LF of LAEs in Section 7.
No correction has been applied for dust.
LAEs are generally recognized to be young, less massive galaxies with little dust ($E(B-V)<0.05$ at $z\sim6$ \citep{ono10} and even at $z\sim$3 \citep{gro07}), though old and massive stellar populations are found in some LAEs \citep{fin09}.

\begin{figure}
\vspace*{-1.4cm}
\hspace*{-0.6cm}
\plotone{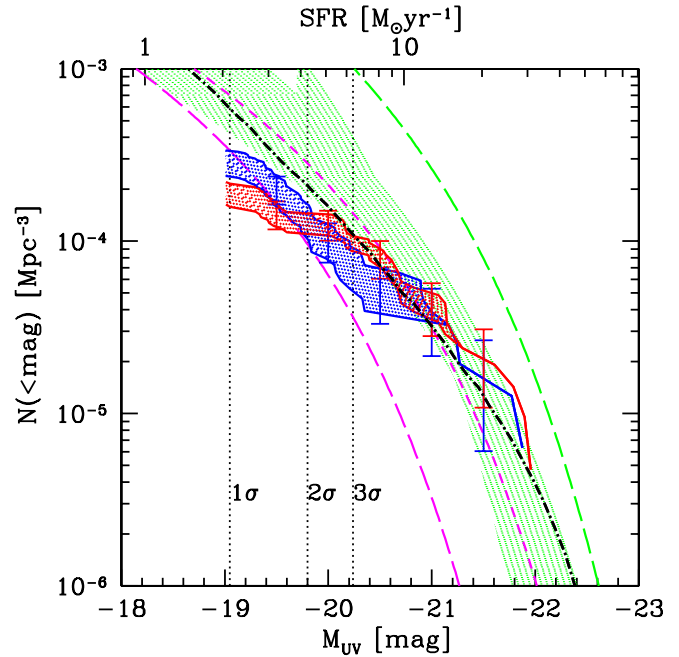}
\epsscale{1.00}
\caption{
Comparison of the rest-UV continuum cumulative LFs of LAEs at $z=6.5$ (red-shaded region) and $z=5.7$ (blue-shaded region).
The upper and lower limits of the shaded regions are determined in the same way as in the Ly$\alpha$ LF estimate. 
The vertical lines indicate the limiting magnitudes in the $z'$ band at $M_{UV}=-19.05$, $-19.80$ and $-20.24$ for $1\sigma$, $2\sigma$, and $3\sigma$ given EW$_0=0$ at $z=6.5$, respectively.
The rest-UV LF measurements at magnitudes fainter than the $3\sigma$ limiting magnitude may be uncertain due to incompleteness in the $z'$ band; we show the rest-UV continuum LFs down to the $1\sigma$ limiting magnitude just for reference.
We did not use data points at $>M_{UV}=-20.24$ when fitting the Schechter function.
Error bars are calculated from Poisson errors.
As a comparison, the rest-UV LF of the LAE sample at $z=5.7$ (magenta short-dashed line) and $z=3.1$ (magenta long-dashed line) evaluated by \citet{ouc08}, the range of the rest-UV LF of the LBG sample at $z\sim6$ (green shaded region) determined by various studies (see Figure 11 of \citealp{bou07}), and the rest-UV LF of $z\sim3$ LBG (\citealp{red08}; green long-dashed line) are shown.
The black dot-dashed line shows the estimate of the UV LF at $z=5.7$ using Ly$\alpha$ LF with $\alpha=-1.5$, assuming a universal EW$_0$-UV luminosity relation (see Section 7).
\label{fig_uvlf}}
\end{figure}

\begin{figure}
\vspace*{-1.2cm}
\hspace*{-0.2cm}
\plotone{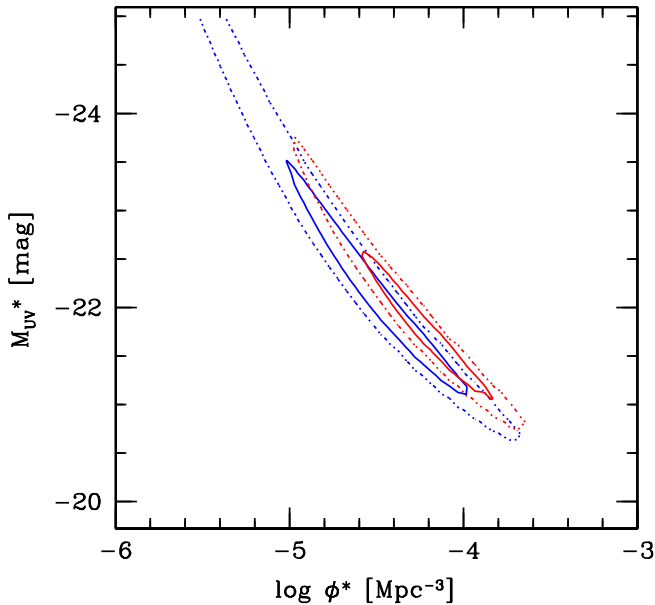}
\epsscale{1.0}
\caption{ 
Error ellipses of the best-fit Schechter parameters $\phi^*$ and $L^*$ of the rest-UV continuum LFs of LAEs at $z=6.5$ (red upper ellipse) and $z=5.7$ (blue lower ellipse), assuming a fixed $\alpha=-1.5$.
The inner and outer solid ellipses are the $1\sigma$ and $3\sigma$ confidence levels, respectively.
The confidence levels of the fit were computed based on the larger of the Poisson error or the difference between the upper limit and the lower limit.
\label{fig_eelip_uv}}
\end{figure}

Figure~\ref{fig_uvlf} shows the rest-UV continuum LF of our LAE sample.
As in the Ly$\alpha$ LF, the red- and blue-shaded regions mark the acceptable UV LF ranges for $z=6.5$ and $5.7$, respectively, and Poisson errors are shown for some data points.
The vertical dotted lines indicate the corresponding limiting magnitudes in the $z'$ band.
Our LF measurements at magnitudes fainter than $M_{UV}=-20.24$ ($3$ $\sigma$) may be uncertain because the corresponding $z'$-band magnitudes are no longer reliable.
The derived rest-UV LFs presented in this study are nearly consistent with our previous studies discussed in K06 and S06, though the amplitudes are slightly lower than before.
We confirmed the result of our previous study, that the rest UV LF does not change between $z=6.5$ and $5.7$, at least at the bright end of $M_{UV}<-20.5$.
We fit the Schechter function to these data points down to $M_{UV}=-20.24$, beyond which our measurements were largely affected by incompleteness of source detection.
Because data points were limited to the bright end, we used a fixed slope of $\alpha=-1.5$.
The best fit parameters were $M_{UV}^*=-21.720^{+0.625}_{-0.875}$ and ${\rm log}(\phi^*)=-4.15^{+0.35}_{-0.40}$ for $z=6.5$, and $M_{UV}^*=-21.845^{+0.625}_{-1.750}$ and ${\rm log}(\phi^*)=-4.30^{+0.35}_{-0.70}$ for $z=5.7$.
Figure~\ref{fig_eelip_uv} shows the error contours, which overlap at the $1\sigma$ errors.
This is in clear contrast to the difference seen in the Ly$\alpha$ LF.
It should be noted that the rest-UV continuum flux is not sensitive to the neutral IGM.
This result strongly supports the interpretation that the difference in Ly$\alpha$ LF between $z=6.5$ and $5.7$ is caused by the IGM attenuation.
However, the best-fit parameters of rest-UV LFs include large uncertainties, and the large error contours in Figure~\ref{fig_eelip_uv} suggest that our LAE sample is still insufficient to strongly constrain the parameters.
Possible distortions in the rest-UV LF caused by photometric error were also evaluated by the Monte Carlo simulation as in Section 4.2.
Assuming $\alpha=-1.5$, the $1\sigma$ fluctuations were found to be $\sigma(M_{UV}^*)=0.253$ and $\sigma( {\rm log}(\phi^*))=0.136$ for $z=6.5$ and $\sigma(M_{UV}^*)=0.154$ and $\sigma( {\rm log}(\phi^*))=0.087$ for $z=5.7$.
These fluctuations are smaller than the fitting errors of the best-fit Schechter parameters, though larger than those of Ly$\alpha$ LFs.

\epsscale{1.4}

\epsscale{1.35}

In Figure~\ref{fig_uvlf}, we overplotted another rest-UV continuum LF estimate (with a simple extrapolation at the faint end) of the LAE sample at $z=5.7$ by \citet{ouc08}, which was almost consistent with our measurement at $z=5.7$ at $M_{UV}<-20.5$.
The green-shaded region is the range of the UV LF of LBGs sampled at $z\sim6$ determined by various studies, based on \citet{bou07}, Figure 11.
The coincidence of UV LFs of LBG and LAE at $z\sim6$ was confirmed by S06 and \citealp{ouc08}.
This indicates that most LBGs at $z\sim6$ display Ly$\alpha$ emissions, though the UV LF measurements remain uncertain for both in LAE and LBG.
It is quite interesting to note that the UV LFs of these two populations are consistent at $z\sim6$ because they generally seem to show different evolutionary trends at $z<6$.
In Figure~\ref{fig_uvlf}, we also present the UV LF of LAEs at $z=3.1$ (\citet{ouc08}; magenta long-dashed line) and that of LBGs at $z\sim3$ (\citet{red08}: green long-dashed line).
The UV LF of the LAEs seems to show an increase in $L^*$ at higher $z$, whereas that of the LBGs seems to have the reverse trend, showing a decrease in $L^*$.
A high fraction of LAEs among LBGs in the early universe is naturally expected if a LBG appears as a LAE during its initial starburst phase, when it is still dust-free \citep{sha01}.
The scenario is supported by measurements of the LAE dark halo mass of $\sim10^{11\pm1}M_\odot$, which is systematically smaller than that of LBGs \citep{ouc10} at all epochs.
\citet{shi06} and \citet{ouc08} reached almost the same conclusion concerning a high fraction of LAEs among LBGs at $z\sim6$, whereas \citet{dow07} claimed that the fraction of LAEs among LBGs at $z\sim6$ was $\sim30\%$, which was almost the same as at $z\sim3$.
\citet{sta11} estimated that the LAE fraction of luminous LBGs with $-21.75<M_{UV}<-20.25$ at $z\sim6$ was as small as $20\%$, though their sample was restricted to strong Ly$\alpha$ emissions with EW$_0>25$\AA.
A much higher fraction of LAEs should be expected at smaller EW$_0$ from their EW distribution.
\citet{hen10} suggested another possibility based on their blind multislit spectroscopic search for LAEs: that the $i$-dropout color selection might miss a certain fraction of LAEs that have blue ($i-z$) colors due to strong Ly$\alpha$ emission.
Otherwise, the UV LF of LBGs might be underestimated due to dust extinction, which is generally small for LAEs.
We cannot make any conclusive argument about the LAE fraction of LBGs and its evolution solely from our observed rest-UV LF; however, one should be careful when comparing among different observations because the fraction depends on the threshold EW applied when selecting the LAEs, as well as on the rest-UV luminosity.




\section{EW distribution}

A more straightforward observable probe of reionization is the equivalent-width distribution of LAEs, which can be derived from the Ly$\alpha$ and UV continuum fluxes.
These are sensitive and insensitive to the neutral IGM, respectively.
It was very hard to measure the rest-frame equivalent width (EW$_0$) in spectroscopic data because most LAEs were too faint to accurately measure their continuum flux directly from spectra.
Instead, the EW$_0$ was calculated using narrow- and broad-band photometry.
EW$_0$ was photometrically estimated in some previous work, though most of the EW$_0$ values were only lower limits because the continuum emission was not detected in broad-band images.
In contrast, most LAEs in the present study were actually detected in the $z'$-band image by virtue of the extreme depth of the SDF images.
The EW$_0$ were reliably determined in those cases.

\begin{figure}
\vspace*{-2.1cm}
\hspace*{-0.7cm}
\plotone{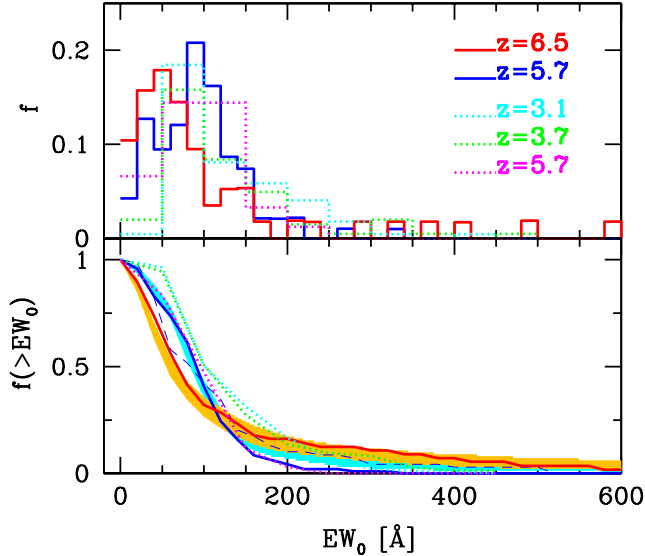}
\epsscale{1.0}
\caption{
Top: the differential EW$_0$ fraction distribution of the LAE sample at $z=6.5$ (red line) and $z=5.7$ (blue line). 
Comparisons with low-$z$ LAEs at $z=3.1$, $3.7$, and $5.7$ from \citet{ouc08} are also shown.
Bottom: the cumulative EW$_0$ fraction distribution of the LAE sample at $z=6.5$ (red line) and $z=5.7$ (blue line). 
The orange and cyan shaded regions are shown as $1\sigma$ fluctuations in the EW$_0$ distribution caused by photometric errors and redshift uncertainties.
The blue dashed line indicates the cumulative EW$_0$ fraction distribution of the LAE sample at $z=5.7$ given that EW$_0$ was calculated from $i^\prime$ and NB816 magnitudes.
\label{fig_ew}}
\end{figure}

Figure~\ref{fig_ew} compares the EW$_0$ distribution of the LAE samples for $z=6.5$ and $5.7$.
The detection completeness was corrected by number weighting according to the NB magnitude, as outlined in Section 4.2.
Here, we did not correct for the absorption of the blue side of the Ly$\alpha$ emission due primarily to the interstellar medium (ISM) absorption inside the galaxy.
K06 showed that the blue-side line profile of Ly$\alpha$ emission of the composite spectrum of LAEs at $z=6.5$ was simply explained by spectral broadening, which meant that the blue side of the observed Ly$\alpha$ emission was almost completely absorbed.
Therefore, if the ISM absorption is corrected, the evaluated EW$_0$ would almost double.
For several LAEs that remained undetected in the $z'$ band ($<1\sigma$), we used the lower limit of EW$_0$ by replacing the $z'$-band magnitude with the $1\sigma$ limiting magnitude of $z'=27.79$.
The systematic errors caused by the process will be discussed later.

In Figure~\ref{fig_ew}, we also plotted the EW$_0$ distribution of lower-$z$ LAEs at $z=3.1$, $3.7$, and $5.7$ extracted from \citet{ouc08}.
The EW$_0$ distribution of our $z=5.7$ LAE sample was almost consistent with other low-$z$ samples, whereas the EW$_0$ at $z=6.5$ seemed to be systematically smaller than that at $z=5.7$.
This appeared more clearly in the cumulative distribution of EW$_0$ shown in the bottom panel of Figure~\ref{fig_ew}.
We evaluated a possible distortion in the EW$_0$ distribution caused by photometric error and the redshift ambiguity by the Monte Carlo simulation, as discussed in Section 4.2.
The $1\sigma$ fluctuations of the EW$_0$ distribution of the simulation, shown as the shaded regions in the bottom panel of Figure~\ref{fig_ew}, ensured that our results were not seriously affected by these uncertainties.
S06 suggested that an estimate of the EW$_0$ distribution at $z=5.7$ based on $i^\prime$ and NB816 magnitudes was largely different from one based on $z^\prime$ and NB816 magnitudes.
In the former estimate, the Ly$\alpha$ line enters both the $i^\prime$ and NB816 bands, as it does at $z=6.5$, which is based on $z^\prime$ and NB921 magnitudes.
The blue dashed line in the bottom panel of Figure~\ref{fig_ew} indicates the EW$_0$ distribution at $z=5.7$ when EW$_0$ is calculated from $i^\prime$ and NB816 magnitudes.
It approaches the EW$_0$ distribution at $z=6.5$, with an extended tail at large EW$_0$.
This might be caused by the fact that the deeper limiting magnitude of the $i^\prime$ band imposed a stronger constraint on the continuum flux than did the $z^\prime$-band, which only provided a lower limit on EW$_0$ when undetected; therefore, deeper BB magnitudes generally tended to yield higher EW$_0$.
Excluding objects that were not detected in the $z^\prime$-band actually reduced the number of objects with high EW$_0$, but did not make a substantial change in the overall EW$_0$ distributions shown in Figure~\ref{fig_ew}, {\it i.e.}, EW$_0$ at $z=6.5$ seems to be systematically smaller than at $z=5.7$.
However, there is no way to estimate the EW$_0$ distribution at $z=6.5$ in the case of using a BB band, in which the Ly$\alpha$ line does not enter, because we do not have any sufficiently deep BB bands at longer wavelength than the $z^\prime$-band.
We cannot completely rule out the possibility that the difference seen in the EW$_0$ distribution between $z=5.7$ and $6.5$ is attributed to the difference whether the Ly$\alpha$ line enters into the BB band or not.
A very deep J-band photometry to estimate EW$_0$ at $z=6.5$ is required to make a more fair comparison of the EW$_0$ distribution.

\epsscale{1.4}

\epsscale{1.4}
\begin{figure}
\vspace*{-2.1cm}
\hspace*{-0.6cm}
\plotone{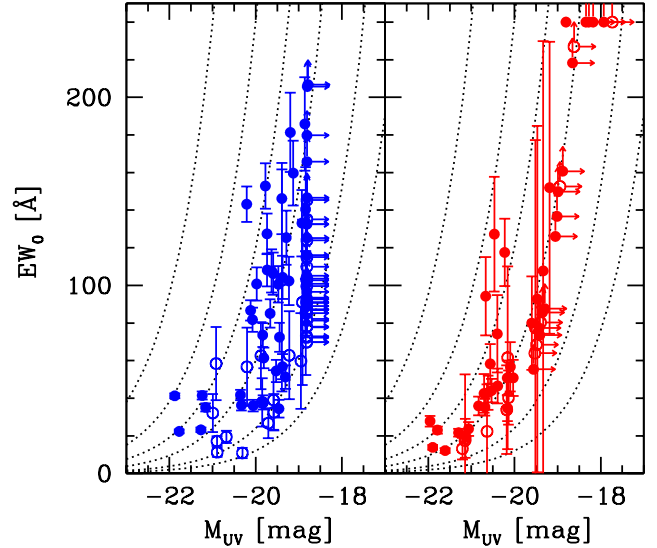}
\epsscale{1.0}
\caption{
UV luminosity-EW$_0$ relation at $z=5.7$ (left) and $6.5$(right).
Filled circles present the spectroscopic sample, and open circles show the photometric sample. 
The objects with EW$_0>$240~\AA~are plotted at the EW$_0=$240~\AA.
Error bars denote uncertainties caused by photometric errors and redshift ambiguity, except for data points that have large uncertainties due to no detection in the BB band (only the upper limit of $M_{UV}$ and the lower limit of EW$_0$ are shown as arrows). 
The dotted curves represent the EW$_0$ at fixed Ly$\alpha$ luminosity; from top to bottom, 5,2,1,0.5,0.2,0.1 $\times10^{43}$ erg s$^{-1}$.
\label{fig_ewuv}}
\end{figure}

The median EW$_0$ at $z=6.5$ is $74$\AA~, which is smaller by $15$\AA~than that at $z=5.7$.
This trend may be easily understood from the resultant LFs of these two epochs. 
The UV continuum LFs are almost identical, whereas Ly$\alpha$ LF at $z=6.5$ is deficient compared with that at $z=5.7$.
The trend was also indirectly suggested from a possible FWHM difference in lines seen by \citet{hu10}, for which it was difficult to derive EW$_0$ due to the shallow photometric data.
\citet{fon10} suggested that the absence of prominent L$\alpha$ lines in their LBG sample at $z\sim7$ compared with a conservative EW distribution at lower-$z$.

Another interesting feature of this plot is that the EW$_0$ distribution at $z=6.5$ has a remarkable extended tail toward larger EW$_0$ compared with $z=5.7$.
Such an effect was also found in the LBG sample \citep{sta07}.
One explanation for this might be contamination by Population III-dominated galaxies.
It is generally thought that the universe was first metal-enriched by this first generation of stars.
As these Population III stars were born in extreme metal-free conditions, they presumably had a top-heavy initial mass function.
Given their exceptionally high effective temperatures, they are expected to produce a very hard spectrum, enough to ionize the primordial He gas.
The main characteristics of the predicted SED are the presence of a large-EW Ly$\alpha$ emission line, due to the strong ionizing flux, and strong He {\sc i} and He {\sc ii} recombination lines, due to spectral hardness \citep{sch02}.
Therefore, Population III-dominated galaxies are expected to appear among high-$z$ LAE samples with large EWs.
Several LAEs at $z=6.5$ have EW$_0$ larger than $\sim300$~\AA, which cannot be attained by the usual Population II synthesis.
These objects are plausible Population III candidates because of their extraordinarily large EW at high $z$, though they are expected to dominate at higher $z$ ($7<z<15$), according to most model predictions (\citealp{joh08}, \citealp{yos07}).
One alternative possible origin for the large Ly$\alpha$ EW is a contribution from AGNs.
Otherwise, the large Ly$\alpha$ EW might be the result of scattering in a clumpy, dusty interstellar medium (\citealp{neu91}; \citealp{han06}; \citealp{fin09}).
Follow-up NIR spectroscopy to detect the HeII emission signal is the only promising way to confirm  Population III-dominated galaxies.
However, it should be noted that such an extended tail toward larger EW$_0$ can, more or less, be produced artificially by the uncertainty in BB flux.

Figure~\ref{fig_ewuv} shows the EW$_0$-UV luminosity relation.
For faint $z'$-band magnitudes ($<27.79$; $1\sigma$), both EW$_0$ and M$_{UV}$ are only provided as lower limits, indicated as arrows in the figure.
Note that the $z'$ bandpass directly corresponds to the UV continuum luminosity at $z=5.7$, whereas the Ly$\alpha$ flux also contributes to the $z'$-band flux at $z=6.5$.
A clear vertical lower-limit sequence of $M_{UV}$ can be seen at $z=5.7$ in Figure~\ref{fig_ewuv}, and not at $z=6.5$.
As reported in previous lower-$z$ studies (S06; \citealp{sta10}; \citealp{van09}; \citealp{ouc08}; \citealp{and06}), we saw an apparent deficit of LAEs with large EW$_0$ at bright UV magnitudes, and the maximum EW$_0$ increased with lower UV luminosity, which is expected if low-luminosity galaxies are less obscured by dust.
This study confirmed the trend at $z=5.7$, and almost the same tendency was found at $z=6.5$, as well.

\section{Ly$\alpha$ Profile of the Composite Spectrum}

The Ly$\alpha$ emission-line profile is also a useful reionization signature, in principle, because the neutral IGM imposes a damping absorption feature on it (\citealp{day08}, \citealp{dij07a}).
A difference in Ly$\alpha$ line profiles between two epochs around the reionization period, might suggest changing IGM opacity.
Though there are many model predictions of Ly$\alpha$ profiles during the reionization epoch, numerous factors such as star formation rate (SFR), internal kinematics, inflow/outflow, and source clustering easily affect the profiles.
On the observational side, individual spectra are usually too noisy to verify the line profile, so several spectra must be stacked to obtain good average line-profile features.
As in our previous study (K06), the composite spectrum was made using the procedure outlined below.
We now have $45$ and $54$ LAE spectra at $z=6.5$ and $5.7$, respectively, though at different spectroscopic resolutions.
First, we removed the spectra with the poorest instrument resolution.
Then, each spectrum was smoothed with a Gaussian kernel chosen to produce a common instrument resolution of FWHM$\sim6.4$ \AA, which was practically measured from the FWHM of sky lines near the Ly$\alpha$ emission for $z=5.7$ and $6.5$, respectively.
Each spectrum was shifted so that the line peak wavelength, which is the only feature used to estimate the redshift, was at the rest wavelength of 1215.67\AA.
The spectra were rebinned to a common pixel scale, and then coadded by taking the average with scaling and weighting based on their line flux, using a $3\sigma$ clipping to eliminate sky-subtraction residuals.

\begin{figure}
\vspace*{-1.2cm}
\hspace*{-0.7cm}
\plotone{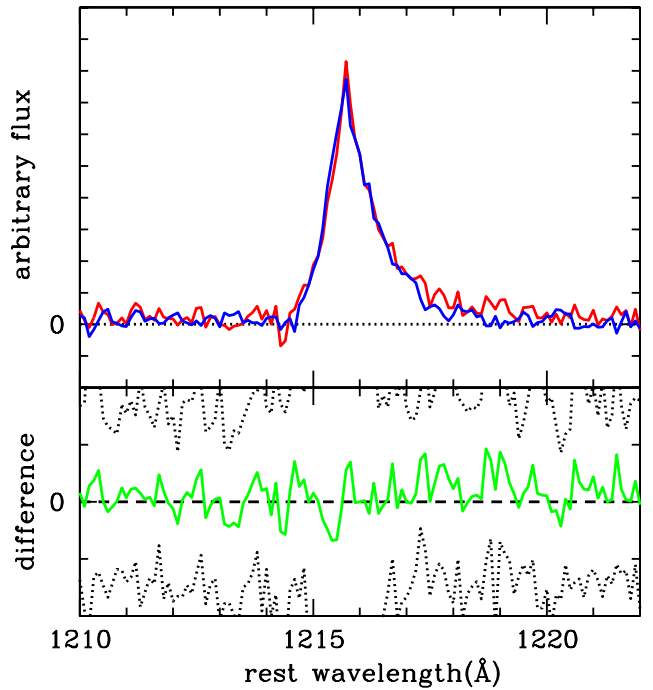}
\epsscale{1.0}
\caption{
Comparison of the composite spectrum between $z=6.5$ (red) and $5.7$ (blue).
The lower panel shows the residual of subtraction of $z=5.7$ from $6.5$.
The dotted line shows the $1\sigma$ error of the difference of the composite spectra.
\label{fig_prof}}
\end{figure}

Figure~\ref{fig_prof} compares the final composite spectrum between $z=6.5$ and $5.7$.
Both composite spectra revealed an apparently asymmetric profile with an extended red wing.
We found no significant differences between these two composite spectra, even when comparing the composite spectra made only from Ly$\alpha$-bright or Ly$\alpha$-faint objects.
This conclusion is consistent with \citet{ouc10} and \citet{hu10}.
The composite spectrum at $z=6.5$ seems to have a slight excess over $z=5.7$ at the red wing tail from $1217$\AA~to $1220$\AA, though it is much smaller than the $1\sigma$ error of the difference in the composite spectra.
This could be caused by a slight broadening in the Ly$\alpha$ emission-line profile itself from $z=5.7$ to $6.5$, as suggested by \citet{ouc10}.
Otherwise, this could be caused by scaling to systematically smaller Ly$\alpha$ flux at $z=6.5$, compared to that of $z=5.7$ (as seen on the EW$_0$ distribution), which increases the UV continuum level at $z=6.5$.

\epsscale{1.45}

The similarity in Ly$\alpha$ emission profiles might indicate a lack of significant IGM opacity evolution from $z=5.7$ to $6.5$.
However, using the composite spectrum of Ly$\alpha$ emission to constrain the reionization has a critical problem of its own.
As mentioned above, the wavelength of the Ly$\alpha$ line peak has to be used to measure the redshift, and is often systematically offset from the rest frame.
\citet{sha03} found the kinematic offset implied by the relative redshifts of Ly$\alpha$ emission and low-ionization interstellar absorption lines in an LBG sample at $z\sim3$.
The offsets vary significantly as a function of Ly$\alpha$ emission strength from $800$ km s$^{-1}$ to $480$ km s$^{-1}$, though these shifts could be smaller at high $z$, where the emitting halos are substantially less massive, and might not power such strong winds as at $z\sim3$.
This can be explained in terms of the properties of large-scale outflows (\citealp{ade03}, \citealp{mas03}, \citealp{wes05}).
Recently, \citet{mcl11} were the first to detect the $[$O {\sc iii}$]$ emission line from two LAEs at $z\sim3$, and they also found apparent velocity offsets.
Such systematic offsets in Ly$\alpha$ emission lines from the systemic redshift would increase the uncertainty in measuring redshifts using only the Ly$\alpha$ peak.
Consequently, when making a composite spectrum using several LAE spectra, these systematic offsets would dilute the line profile, critically preventing us from revealing an accurate line profile.
Also, if the field-to-field variances of the neutral fraction and the size of H {\sc ii} bubble were large even at the same redshift, the composite line profile would be diluted.
It seems hard to derive a conclusive constraint on reionization from line profiles alone.
The most promising method for overcoming the problem is to measure the systemic redshift of LAEs by detecting nebular emissions (\citealp{ade05}, \citealp{mcl11}), which is still technically difficult at $z>5$.
This wind effect could allow the Ly$\alpha$ emission line to emerge at wavelengths where the GP optical depth is reduced, transmitting the Ly$\alpha$ flux directory to the observer even at the reionization epoch \citep{dij10}.
\citet{mcq07} found that a $400$ km s$^{-1}$ redshift in the Ly$\alpha$ line did not have a large effect on their conclusions regarding the effect of reionization on the Ly$\alpha$ LF and clustering at $x_{\rm H I}<0.4$, and they concluded that this wind effect would not seriously affect the ^^ ^^ Ly$\alpha$ test" in the late phase of reionization.

We would also like to note that the Ly$\alpha$ radiative transfer is complicated by the geometry and kinematics of the ISM and IGM.
The relative geometries of interstellar H {\sc i} and H {\sc ii} regions significantly affect resonant scattering, which can either suppress or enhance the Ly$\alpha$ line (\citealp{cha93}, \citealp{neu91}).
The resonant scattering in the IGM induces a change in the frequency of Ly$\alpha$ photons, which can also cause a mass-dependent redward shift of the Ly$\alpha$ line peak, even in the absence of a galactic wind \citep{zhe10}.
Dust attenuation, if any, also significantly reduces the Ly$\alpha$ emission.
A stacking analysis would, more or less, overlook these variations among galaxies.

In K06, we showed that our composite Ly$\alpha$ line profile at $z=6.5$ could be realized by both the reionization model, in which we included the attenuation of a GP damping wing from outside the H {\sc ii} bubble, and the galactic wind model, which has another broadly extended Gaussian component in the line profile.
Although additional spectroscopic data have improved the quality of the composite spectrum, the spectral resolution of our composite spectrum is still too low to distinguish between these two models.
Higher resolving power provided by larger telescopes will be required to constrain the model more strongly.

\section{Contribution of LAEs to the reionization photon budget}

The integration of the observed UV LF at the faint end provides an estimate of the luminosity density and, thus, of the photon budget of reionization.
Therefore, an accurate UV LF estimate based on the LAE sample will constrain the LAE contribution to the photon budget.
Unfortunately, the UV LF measurements at magnitudes fainter than $M_{UV}=-20.24$ ($3\sigma$) may be critically uncertain because of incompleteness, which cannot be corrected due to the difference in the completeness of the NB and $z'$-band data.
Here, we take another approach to derive the UV LF of LAEs using both the Ly$\alpha$ LF and the EW$_0$-UV luminosity relation, as shown in Figure~\ref{fig_ewuv}.

The EW distribution function is well approximated by an exponential function (\citealp{gro07}; \citealp{sta11}), though its $e$-folding width, $w$, might depend on UV luminosity.
Several previous works provide derivations of the EW$_0$-UV luminosity relation of LAE at lower-$z$ (e.g., \citealp{sta10}; \citealp{van09}; \citealp{ouc08}; \citealp{and06}).
Their conclusions are roughly consistent with an apparent deficit of large EW$_0$ at bright UV magnitudes and a maximum EW$_0$ that increases with lower UV luminosity.
The relation does not show a clear evolution across the $3<z<6$ redshift range (\citealp{ouc08}, \citealp{cas11}), and our result at $z=6.5$ also roughly follows this trend (Figure~\ref{fig_ewuv}), though we derived our result from a relatively small sample.
Among these previous studies, \citet{sta10} derived a reliable EW$_0$-UV luminosity relation based on a large spectroscopic LAE sample at $3<z<7$.
Based on this relation, the characteristic $e$-folding width, $w$, was empirically determined to increase with fainter UV magnitudes: $w =60M_{UV}+1440$, which gives $w=60$\AA~at $M_{UV}=-23$ and $w=300$\AA~at $M_{UV}=-19$, respectively.
This equation almost traces the maximum EW$_0$ values as a function of M$_{UV}$.
Figure $12$ of \citet{sta10} shows an apparent deficiency in LAEs with low EW$_0$ and low UV luminosity, though it could be caused by their spectroscopic detection limit. 
Hence, we assumed that the peak of the exponential distribution was always at EW$_0=0$, independent of M$_{UV}$.
A cutoff EW, below which LAEs were not included in the sample, was set to $10$\AA~, based on the LAE selection criterion $m_z-m_{NB}>1$, and Equation (1).
With this EW$_0$ probability distribution, we conducted Monte Carlo simulations to derive M$_{UV}$, given a Ly$\alpha$ luminosity and EW$_0$.
The Ly$\alpha$ LF measurements included an accurate correction for completeness; therefore, the UV LF was estimated free of incompleteness even at the faint end when assuming the EW$_0$ distribution, as described above.

The dot-dashed line in Figure~\ref{fig_uvlf} shows the estimate of the UV LF at $z=5.7$ with this method.
Here, we assumed that the faint end of the Ly$\alpha$ LF at $z=5.7$ was $\alpha=-1.5$.
Note that we did not use the UV LF at $z=6.5$ to constrain the photon budget because the Ly$\alpha$ LF at $z=6.5$ might be affected by neutral IGM attenuation.
The UV LF estimated with the method is fairly consistent with the bright end of the observed UV LF at $z=5.7$, where the incompleteness is not severe.
The bright end of the rest-UV LF down to $3\sigma$ limiting magnitudes in the $z'$ band can reasonably be reproduced from the observed Ly$\alpha$ LF and the EW$_0$-UV luminosity relation.
However, the shape of the derived UV LF is closer to a power law than to a Schechter function.
When we fit the UV LF to a Schechter function, the faint-end slope was found to be $\alpha=-2.4$, which is much steeper than those derived from LBGs \citep{bou07}.

\begin{figure}
\vspace*{-1.2cm}
\hspace*{-0.6cm}
\plotone{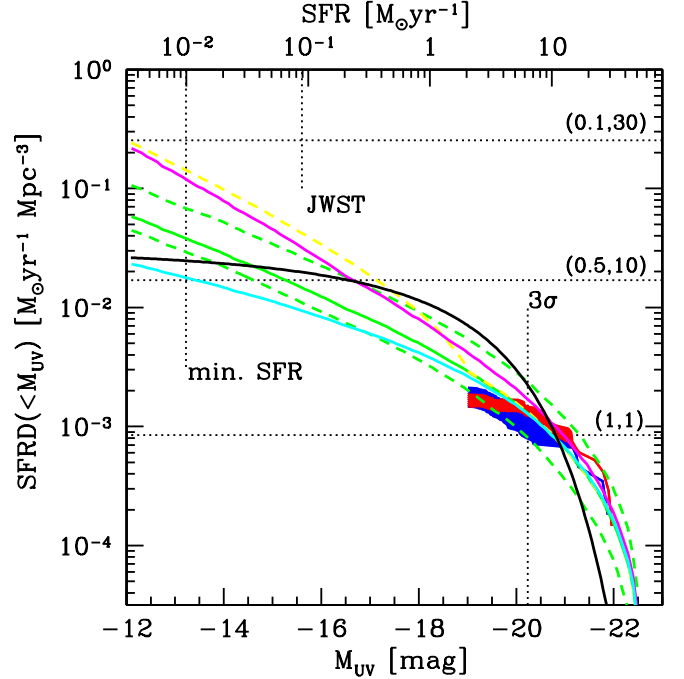}
\epsscale{1.0}
\caption{
Comparison between $\dot{\rho}_*$ for different parameter sets of ($f_{\rm esc}$, $C$) in Equation (2) and the SFRD (green solid line) derived by integrating the UV LF of LAEs at $z=5.7$ estimated in Section 7.
The SFRD estimates based on the observed UV LF appear as the blue (red) shaded region for $z=5.7$ ($6.5$), though these are incomplete at the faint end below the $3\sigma$ limiting magnitude.
The dashed line above the solid line shows the SFRD estimates when $w$ is reduced to half.
The dashed line below the solid line shows the SFRD estimates when the cutoff EW$_0=15$\AA~instead of $10$\AA is applied.
The magenta and cyan solid lines show the SFRD estimate when $\alpha$ is changed from $-1.5$ to $-1.7$ and $-1.3$, respectively.
The black solid line shows the SFRD estimate based on the UV LF of LBGs at $z\sim6$ \citep{bou07}.
The minimum SFR inferred by the simulation by \citet{nag10} and approximate detection limit of JWST are shown by dotted lines.
\label{fig_sfrd}}
\end{figure}

The critical number density of ionizing photons necessary to keep the intergalactic hydrogen ionized was given in \cite{mad99}.
The corresponding critical star formation rate density (SFRD), $\dot{\rho}_*(z)$, can be written:

\begin{eqnarray}
\dot{\rho}_*(z) = 0.013 f_{\rm esc}^{-1} \left(\frac{C}{30}\right) \left(\frac{1+z}{6}\right)^3 \left(\frac{\Omega_b h_{70}^2}{0.04}\right)^2 M_\odot {\rm yr}^{-1} {\rm Mpc}^{-3},
\end{eqnarray}

\noindent where $f_{\rm esc}$ is the escape fraction of ionizing photons, and $C$ is the ionized hydrogen clumping factor of the IGM.
Here, the authors assumed a Salpeter IMF, solar metallicity, and SED at $<912$\AA~based on the Bruzual \& Charlot population synthesis model.
Both $f_{\rm esc}$ and $C_{30}$ are highly uncertain.
The $f_{\rm esc}$ was generally estimated to be $<0.1$ (\citealp{lei95}, \citealp{deh01}, \citealp{fer03}, \citealp{mal03}, \citealp{sha06}) for various galaxies at various redshifts.
\citet{iwa09} and \citet{ino11} obtained much higher $f_{\rm esc}$ of $\sim0.5$ for some LAEs at $z=3.1$, and \citet{van10} found one LBG at $z=3.795$ with a direct detection of the Lyman continuum flux, which gave $f_{\rm esc}>0.15$. 
\citet{bou10a} recently argued for the possibility of high $f_{\rm esc}$ at $z\sim7$ to account for the observed steep UV-continuum slope without a nebular emission contribution (see also \citealp{tan10}).
We assumed that the escape fraction did not depend either on luminosity or on redshift; though some models predict that it could increase up to $\sim0.8$ in lower-mass galaxies at higher $z$ (\citealp{ric00}, \citealp{wis09}, \citealp{raz10}).
\citet{sia10}, based on deep HST far-UV images, suggest that the escape fraction significantly increases from $z=1.3$ to $3$.
The IGM clumping factor was estimated to be $C\sim30$ at $z=5$ based on a numerical simulation, though it could be lower if higher density regions are less ionized \citep{gne97}.
A more recent simulation by \citet{paw09} suggests a much lower clumping factor, $C=6$.
\citet{bh07} suggest $C \lesssim3$ at $z=6$, based on measurements of the metagalactic photoionization rate, combined with a model for the ionizing photon mean free path.
Figure~\ref{fig_sfrd} presents a comparison between $\dot{\rho}_*$ for different parameter sets of ($f_{\rm esc}$, $C$) in Equation (2) and the SFRD derived by integrating the UV LF estimated in this section.
Assuming the complete escape of ionizing photons ($f_{\rm esc}=1$) and a homogeneous IGM ($C=1$), the universe can be easily ionized only by bright LAEs at $z=6.5$; however, in the more realistic case with ($f_{\rm esc}<1$ and $C \gg1$), the contribution of low-luminosity LAEs is important to the reionization process.

\epsscale{1.45}

Our method assumed the EW$_0$-UV luminosity relation, which is found to be quite sensitive to the SFRD estimate.
The dashed line above the solid line in Figure~\ref{fig_sfrd} shows the SFRD estimates when $w$ was reduced to half.
The dashed line below the solid line in Figure~\ref{fig_sfrd} shows the SFRD estimates when the EW$_0=15$\AA~cutoff was applied instead of $10$\AA.
Both of these estimates are apparently inconsistent with the reliably observed bright end of the SFRD; therefore, only narrow ranges of parameters in the EW$_0$-UV luminosity relation are acceptable.
As mentioned in Section 5, an EW$_0$ larger than $\sim300$\AA~is hard to explain using the usual Population II synthesis.
When making a trial cutoff at the high-EW$_0$ end of the EW$_0$-UV luminosity, the SFRD steepens sharply at the faint end.
We also fit a one-sided Gaussian instead of an exponential to the EW$_0$ distribution function, but the estimate did not change significantly.
The most uncertain parameter in the estimate was the faint-end slope of the Ly$\alpha$ LF, $\alpha$, which, thus far, was poorly constrained by observation.
The magenta and cyan solid lines in Figure~\ref{fig_sfrd} show the SFRD estimates when changing $\alpha$ from $-1.5$ to $-1.7$ and $-1.3$, respectively.
Although these two estimates hardly affect the bright end, maintaining consistency with the observations, they diverge strongly at the faint end.
The steeper faint-end slope makes a larger contribution to the photon budget, as expected.

The black solid line in Figure~\ref{fig_sfrd} shows the SFRD estimate based on the UV LF of LBGs at $z\sim6$ \citep{bou07}.
Although the SFRD of LAEs seems to be larger than that of LBGs at the bright end, they are within the uncertainties (see the shaded region of Figure~\ref{fig_uvlf} indicating the range of the UV LF of the LBG sample at $z\sim6$ determined by various studies).
At the faint end, the SFRD of LAEs increases more significantly than the SFRD of LBGs toward faint magnitudes, as expected from the fact that the LAEs have steeper UV LF. 
It should be noted that this result strongly depends on the faint-end slope, $\alpha$, of the Ly$\alpha$ LF at $z=5.7$, which is poorly constrained by this study.
Here, we provide only estimates for values of $\alpha=-1.3$, $-1.5$ and $-1.7$.
At low-$z$, \citet{gro07} derived a faint-end slope of Ly$\alpha$ LF of $\alpha=-1.49^{+0.45}_{-0.34}$, and \citet{cas11} found $\alpha=-1.60^{+0.12}_{-0.12}$ at $z\sim2.5$ and $\alpha=-1.78^{+0.10}_{-0.12}$ at $z\sim4$.
It is interesting that a steeper faint-end slope of the UV LF than of the LBG is predicted in Figure~\ref{fig_sfrd} even when $\alpha=-1.3$, which is shallower than those at low $z$.
\citet{sta11} also suggested a higher faction of LAEs in their LBG sample towards fainter UV magnitudes.
Consequently, the relative contribution to the photon budget of the LAEs compared with the LBGs increases toward fainter magnitudes.
It finally exceeds the contribution of the LBGs at the faintest magnitude, which seems unlikely.
\citet{nag10} employed a cosmological SPH simulation to reproduce some observational properties of LAEs and found a critical threshold in stellar mass around $M_*\sim10^7M_\odot$, below which star formation rapidly drops, as might be expected from the Kennicutt law.
Therefore, we assumed that the minimum SFR that galaxies should have is $10^{-2}M_\odot$ yr$^{-1}$, which corresponds to the critical stellar mass in the simulation by \citet{nag10}.
When assuming $\alpha=-1.5$, the LAE contribution exceeds the LBG around $M_{UV}\sim-15$. 
This might suggest that some of our assumptions are not reasonable; otherwise, the faint-end slope of Ly$\alpha$ LF of LAEs should be shallower than $-1.5$.
For example, given $\alpha=-1.3$, the LAE contribution does not overwhelm the LBG contribution down to SFR$=10^{-2} M_\odot$ yr$^{-1}$.
However, we note here again that the derived UV LF of LAEs is based on many assumptions.
For example, the faint-end slope of the SFRD becomes flatter even with $\alpha=-1.5$ if much larger EW$_0$ are allowed at faint UV luminosities.
On the other hand, we note that the LBG estimate is also based on an extrapolation of the observed UV LF to very faint levels.

Despite many uncertainties, this might be the first estimate of the contribution of the LAE population to the reionizing photons.
The SFR estimate based on Ly$\alpha$ luminosity is much more strongly affected by extinctions of dust, ISM, and IGM than the SFR estimate based on rest-UV luminosity.
Our result indicates that low-luminosity LAEs could contribute significantly to the photon budget necessary for reionization, though these faint LAEs are below the current detection limits.
Insofar as a universal EW$_0$-UV luminosity relation is assumed, a steep faint-end slope in the UV LF is predicted even when the faint-end slope of the Ly$\alpha$ LF is shallower than those at low $z$.
The faint end of the LF of these ionizing sources is critical to the conclusion, and an accurate determination of the faint end slope of Ly$\alpha$ LF will enable a precise estimate of the total ionizing photon density emitted by LAEs at this epoch.
\citet{bou07} concluded that the number of $i$ dropouts at $z\sim6$ appears to be approximately consistent with the numbers necessary to reionize the universe.
Recent studies using HST/WFC3 identified more and fainter galaxies at even higher $z$, suggesting a faint-end slope that is sufficiently steep (\citealp{oes10}, \citealp{bou10b}) to fully reionize the universe.
The LAE sample based on NB searches exploring only a small redshift coverage is complementary to the estimate because LAEs should constitute a very young population at a single epoch among heterogeneous high-$z$ star-forming galaxies sampled by the $i$-dropout method.
Moreover, our LAE sample has the advantage of many spectroscopic confirmations, providing an accurate estimate of sample completeness and contamination, which also increases the reliability of our photometric LAE sample.



\section{Summary and Discussion}

We obtained extended spectroscopic confirmations of LAEs at $z=6.5$ and $5.7$ in the SDF, and our conclusions can be summarized as follows:

1. We provided new identifications of $28$ and $20$ LAEs at $z=6.5$ and $5.7$, respectively. 
The discrimination of an LAE from other nearby emitters was based on a quantitative line asymmetric estimator, weighted skewness, $S_w$.
The total number of spectroscopically confirmed LAEs in the SDF is now $45$ ($54$) at $z=6.5$ ($5.7$), which means that $90\%$ ($74\%$) of the photometric candidates have been followed by spectroscopy.
Our long campaign of follow-up spectroscopy shows that our photometric LAE sample is highly reliable, with low incompleteness and little contamination.

2. We made more careful measurements of Ly$\alpha$ luminosity, both photometrically and spectroscopically, than in our previous study to determine Ly$\alpha$ and rest-UV LFs more accurately. 
The non-square transmission curve of filters was taken into account in the photometric measurements, and we corrected for slit loss in the spectroscopic measurements.
The Ly$\alpha$ fluxes measured photometrically and spectroscopically agree very well, showing that our measurement of Ly$\alpha$ luminosity is accurate.

3. We derived Ly$\alpha$ LFs of LAEs at $z=6.5$ and $5.7$.
With a large number of spectroscopic confirmations of our LAE sample and more careful measurements of luminosities, the Ly$\alpha$ LFs at both redshifts are more sharply determined than those in our previous studies.
The substantially improved evaluation of Ly$\alpha$ LF of $z=6.5$ shows an apparent deficit from $z=5.7$, at least at the bright end and a possible decline even at the faint end, though small uncertainties remain.

4. We derived rest-UV LFs of LAEs at $z=6.5$ and $5.7$.
We confirmed the result of our previous study, namely that the rest-UV LFs at $z=6.5$ and $5.7$ agree with each other, which is in clear contrast to the difference seen in the Ly$\alpha$ LF, though the measurements of UV LF of LAEs still have large uncertainties.
The rest-UV LF of LAEs is almost consistent with that of LBGs within the errors, at least at the bright end, suggesting that the fraction of LAEs among LBGs at $z\sim6$ is higher than that at $z=3$.

5. The EW$_0$ distribution of our $z=5.7$ LAE sample is almost consistent with other low-$z$ samples, whereas the EW$_0$ at $z=6.5$ seems to be systematically smaller than $z=5.7$.
This may be understood from the resultant LFs of these two epochs: UV-continuum LFs are almost identical, whereas the Ly$\alpha$ LF at $z=6.5$ is deficient compared to that at $z=5.7$.
There remains, however, the possibility that the trend is attributed to the artifacts of using BB band including the Ly$\alpha$ line at $z=6.5$.
The EW$_0$ distribution at $z=6.5$ shows an extended tail toward larger EW$_0$ compared with $z=5.7$, though EW$_0$ measurements are not reliable when not detected in the BB band.
The EW$_0$ and rest-UV luminosity relation shows an apparent deficit of LAEs with large EW$_0$ at bright UV magnitudes, and the upper-limit of the EW$_0$ increases with lower UV luminosity.

6. We found no significant difference in the composite Ly$\alpha$ line profile between $z=6.5$ and $5.7$, though the possible velocity offset of Ly$\alpha$ line from the systemic redshift might dilute the composite profiles.

7. We tried to recover the rest-UV LF of LAEs at the faint end, where incompleteness is severe, by assuming a universal EW$_0$-UV luminosity relation.
When choosing reasonable parameters to fit to the observed EW$_0$-UV luminosity relation, the bright end of the rest-UV LF at $z=5.7$ was well reproduced from the observed Ly$\alpha$ LF.
Integrating this rest-UV LF permitted the first experimental estimate of the photon budget of LAEs for reionization.
The derived UV LF suggested that the fractional LAE contribution to the photon budget among LBGs significantly increases toward fainter magnitudes.
Low-luminosity LAEs could contribute significantly to the photon budget, though this depends on the poorly constrained faint-end slope of the Ly$\alpha$ LF.

\subsection{Implications for Reionization}

We confirmed our previous result that the Ly$\alpha$ LF at $z=6.5$ declines from those at $z<5.7$ based on our deep LAE samples with a high spectroscopic identification rate.
This result is, more or less, consistent with \citet{ouc10} and \citet{hu10}.
The decline in the Ly$\alpha$ LF of LAEs from $z=5.7$ to $6.5$ could be caused by the evolution of some intrinsic property of LAEs or by evolution in the ionization state of the IGM.
It will be almost impossible to distinguish between these two possibilities from the Ly$\alpha$ LF alone; however, the rest-UV LF, which is not sensitive to the neutral IGM, may provide an additional diagnostic of the cause of the observed trend.
Interestingly, our measurements of the rest-UV LFs agree well within uncertainties between these two epochs at least at the bright end, where an apparent difference is observed in the Ly$\alpha$ LF.
This result might support the interpretation of neutral IGM attenuation.
Assuming a fully ionized IGM at $z=5.7$, the observed difference in the Ly$\alpha$ LF suggests $x_{\rm H I}\sim0.38$ at $z=6.5$ based on the model of \citet{mcq07}.
Another important aspect of ascribing the difference in Ly$\alpha$ LF to the galaxy evolution of LAEs is that it should simultaneously explain the difference between $z=5.7$ and $6.5$ and the lack of evolution in the Ly$\alpha$ LF from $z=3$ to $6$.
The model of \citet{kob10}, which takes into account a possible evolution of $f_{\rm esc}$ and reasonably reproduces Ly$\alpha$ LF, UV LF, and the EW distribution of LAEs from $z=3.1$ to $6.5$, favors $x_{\rm H I}\sim0.4$ at $z=6.5$.
In the model of \citet{kob10}, the observed large decline in the Ly$\alpha$ LF at $z=6.5$ could be partly caused by galaxy evolution; however, IGM attenuation is still required to account for the entire decline.
The predicted value of the neutral fraction at the moment is strongly model-dependent.
\citet{day08} concluded that the Ly$\alpha$ LF at $z=6.5$ could be reproduced by their model even with small $x_{\rm H I}=3\times10^{-4}$, though a strong increase in dust content was required to match the model to the Ly$\alpha$ LF at $z<5$.
The amplitude difference in the Ly$\alpha$ LF is still relatively small between these two epochs, suggesting that the universe was still largely ($\ga 60\%$) ionized at $z=6.5$.
Such an amplitude difference is smaller than the uncertainties of the rest-UV LFs, though we concluded that the rest-UV LFs between the two epochs agree well at the bright end.
More accurate measurement of the rest-UV LFs based on the larger LAE sample in the future will provide a stronger constraint on the Ly$\alpha$ test.

\subsection{Cosmic Variance Uncertainties}

The decline in Ly$\alpha$ LF from $z=5.7$ to $6.5$ might be caused by cosmic variance.
The sample size is not yet large enough to statistically refute this interpretation; however, we found no difference in the rest-UV continuum LFs, which should be affected by cosmic variance from $z=5.7$ to $6.5$, if it exists.
\citet{ouc10} recently carried out an LAE survey at $z=6.5$ with an FOV five times wider than this study.
Though their number of spectroscopic confirmations is small, the result is statistically robust and less affected by cosmic variance than ours.
They concluded that the Ly$\alpha$ LF at $z=6.5$ shows a decline of $30\%$ in Ly$\alpha$ luminosity from $z=5.7$, whereas the present study finds a $24\%$ decline.
Interestingly, they detected an apparent difference in the LF even at the faint end.
Their LF difference alone constrains the neutral fraction to $x_{\rm H I}<0.2^{+0.2}_{-0.2}$, though they did not detect the significant enhancement of clustering amplitude expected during reionization.
\citet{hu10} also completed their LAE surveys at $z=6.5$ and $5.7$, using an FOV five times wider than this study.
Though the shape of their LF differs from this study and \citet{ouc10}, they also found that the Ly$\alpha$ LFs differed based on their own estimates at these two epochs, indicating a $44\%$ decline in $\phi^*$, whereas this study found a $34\%$ decline.
It is interesting that both of these wide-field studies confirmed a difference in the Ly$\alpha$ LF between $z=6.5$ and $5.7$.
\citet{nak11} also derived the Ly$\alpha$ LF at $z=6.5$, with a low number density only $\times0.3$ of our previous study.
The degree of difference in the LF found in the present study, \citet{ouc10}, and \citet{hu10} are slightly inconsistent, possibly suggesting field-to-field cosmic variance.
Otherwise, such a variance could be caused by a patchy reionization process, which has been suggested in the GP trough measurements showing a substantial variation in IGM transmissions among different QSO lines of sight around $z\sim6$ \citep{djo06}.
Such a process is expected to be caused by primordial clustering or an initial large-scale structure of ionizing sources.
Much wider surveys of high-$z$ LAEs are required to obtain statistically significant constraints on any variance.

\subsection{Photon budget of Reionization}

In Section 7, we attempted to estimate the rest-UV LF of LAEs using the Ly$\alpha$ LF, assuming a universal EW$_0$-UV luminosity relation.
Interestingly, this estimate reproduces the bright end of the observed UV LF reasonably well, enabling an estimate of the LAE contribution to the photon budget required for reionization.
We note, however, that this method has large uncertainties, including the universality of the EW$_0$ distribution function and the EW$_0$-UV luminosity relation.
\citet{pen10} indicated that a correlation between Ly$\alpha$ strength and age or SFR might change with cosmic time, potentially changing the EW$_0$ distribution function.
\citet{nil09} suggested no correlation between EW and UV luminosity; however, this assumption, coupled with the observed Ly$\alpha$ LF, produces a much steeper faint-end slope in the UV LF of LAEs, exacerbating the inconsistency with the UV LF of LBG.
The universality of the EW$_0$ distribution and EW$_0$-UV luminosity relation should be verified at lower $z$ based on a larger LAE sample.
The observed large scatter in Ly$\alpha$ EW$_0$ at faint UV luminosity could be due to small amounts of dust extinction \citep{ver08} and possibly to a stochasticity of their complicated duty circle (\citealp{nag10}), which has not yet been clearly revealed.
Furthermore, more accurate determinations of the faint-end slope of the Ly$\alpha$ LF, $f_{\rm esc}$, and $C$ are required to provide a stronger constraint on the photon budget.
The {\it James Webb Space Telescope} (JWST) will extend the current observational limit down to $m_{AB}\sim31$, which corresponds to $M_{UV}=-15.6$ at $z=5.7$ (see Figure~\ref{fig_sfrd}).
Significant detections of LAEs down to this faint end will allow a more precise estimate of the photon budget.
The JWST/NIRCam and TFI will catch LAEs at higher $z$ far beyond the current frontier of the distant universe.
The derived Ly$\alpha$ LF during the early reionization phase will be more sensitive to the neutral fraction, giving a stronger constraint on the history of reionization.

\acknowledgments

We are grateful to the Subaru and Keck Observatory staffs for their help with the observations.
The observing time for part of this project was committed to all the Subaru Telescope builders.
We thank an anonymous referee for helpful comments that improved the manuscript.
This research was supported by the Japan Society for the Promotion of Science through Grant-in-Aid for Scientific Research 19540246.



{\it Facilities:} \facility{Subaru (FOCAS, Suprime-Cam)}, \facility{KeckII (DEIMOS)}.


\begin{thebibliography}{}
\bibitem[Adelberger et al.(2005)]{ade05} Adelberger, K.L. et al. 2005, \apj 629, 639
\bibitem[Adelberger et al.(2003)]{ade03} Adelberger, K.L. et al. 2003, \apj 584, 45
\bibitem[Ando et al.(2006)]{and06} Ando, M., et al. 2006, \apj, 645, L9
\bibitem[Barkana \& Loeb(2004)]{bar04} Barkana, R., \& Loeb, A. 2004, \apj, 609, 474
\bibitem[Bolton \& Haehnelt(2007)]{bh07} Bolton, J.S., \& Haehnelt, M.G. 2007, \mnras, 382, 325
\bibitem[Bouwens et al.(2007)]{bou07} Bouwens, R.J. et al. 2007, \apj, 670, 928
\bibitem[Bouwens et al.(2010a)]{bou10a} Bouwens, R.J. et al. 2010, \apj, 708, 69
\bibitem[Bouwens et al.(2010b)]{bou10b} Bouwens, R.J. et al. 2010, arXiv:1006.4360
\bibitem[Cassata et al.(2011)]{cas11} Cassata, P. et al., 2011, \aap, 525, 143
\bibitem[Charlot \& Fall(1993)]{cha93} Charlot, S., \& Fall, S.M. 1993, \apj, 415, 580
\bibitem[Dayal et al.(2008)]{day08} Dayal, P. et al., 2008, \mnras, 389, 1683
\bibitem[Dayal et al.(2009)]{day09} Dayal, P. et al., 2009, \mnras, 400, 2000
\bibitem[Deharveng et al.(2001)]{deh01} Deharveng, J.-M. et al. 2001, \aap, 375, 805
\bibitem[Dijkstra et al.(2010)]{dij10} Dijkstra, \& M, Wyithe, J.S.B. 2010, \mnras 408, 352
\bibitem[Dijkstra et al.(2007a)]{dij07a} Dijkstra, M, Litz, A., \& Wyithe, J.S.B. 2007, \mnras 377, 1175
\bibitem[Dijkstra et al.(2007b)]{dij07b} Dijkstra, M, Wyithe, J.S.B. \& Haiman, Z., 2007, \mnras, 379, 253
\bibitem[Djorgovski et al.(2006)]{djo06} Djorgovski, S.G., Bogosavljevic, M. \& Mahabal, A., 2006, NewAR, 50, 140
\bibitem[Dow-Hygelund et al. (2007)]{dow07} Dow-Hygelund, C.C. et al. 2007, \apj, 660, 47
\bibitem[Faber et al.(2003)]{fab03} Faber, S. et al. 2003, Proc. SPIE, 4841, 1657
\bibitem[Fan et al.(2006)]{fan06} Fan, X. et al., 2006, \aj, 132, 117
\bibitem[Fernandez-Soto et al.(2003)]{fer03} Fernandez-Soto, A., Lanzetta, K.M., Chen, H.-W. 2003, \mnras, 342, 1215
\bibitem[Finkelstein et al.(2009)]{fin09} Finkelstein, S.L. et al. 2009, \apj, 691, 465
\bibitem[Fontana et al.(2010)]{fon10} Fontana, A. et al. 2010, \apjl, 725, 205
\bibitem[Furlanetto et al.(2004)]{fur04} Furlanetto, S.R., \& Hernquist, L., Zaldarriaga, M. 2004, \mnras, 354, 695
\bibitem[Gnedin \& Ostriker(1997)]{gne97} Gnedin, N.Y. \& Ostriker, J.P. 1997, \apj, 486, 581
\bibitem[Gronwall et al.(2007)]{gro07} Gronwall, C. et al. 2007, \apj, 667, 79
\bibitem[Haiman \& Spaans(1999)]{hai99} Haiman, Z. \& Spaans, M. 1999, \apj, 518, 138
\bibitem[Haiman \& Cen(2005)]{hai05} Haiman, Z. \& Cen, R. 2005, \apj, 623, 627
\bibitem[Henry et al.(2010)]{hen10} Henry, A.L. et al. 2010, \apj, 719, 685
\bibitem[Hansen \& Oh (2006)]{han06} Hansen, M. \& Oh, S.P. 2006, \mnras, 367, 979
\bibitem[Hibon et al.(2010)]{hib10} Hibon, P. et al. 2010, \aap, 515, 97
\bibitem[Hu et al.(2010)]{hu10} Hu, E.M., Cowie, L.L., Barger, A.J., Capak, P., Kakazu, Y., Trouille, L. 2010, \apj, 725, 394
\bibitem[Hu et al.(2002)]{hu02} Hu, E.M., Cowie, L.L., McMahon, R.G., Capak, P., Iwamuro, F., Kneib, J.-P., Maihara, T., Motohara, K. 2002, \apj, 568, L75; Erratum, 576, L99
\bibitem[Iliev et al.(2008)]{ili08} Iliev, I.T. et al. 2008, \mnras, 391, 63
\bibitem[Inoue et al.(2011)]{ino11} Inoue, A.K. et al. 2011, \mnras, 411, 2336
\bibitem[Iwata et al.(2009)]{iwa09} Iwata, I. et al. 2009, \apj, 692, 1287
\bibitem[Iye et al.(2006)]{iye06} Iye, M. et al. 2006, Nature, 443, 186
\bibitem[Jiang et al.(2008)]{jia08} Jiang, L. et al. 2008, \aj, 135, 1057
\bibitem[Johnson et al.(2008)]{joh08} Johnson, J.L. et al. 2008, \mnras, 388, 26
\bibitem[Kashikawa et al.(2002)]{kas02} Kashikawa, N. et al. 2002, \pasj, 54, 819
\bibitem[Kashikawa et al.(2004)]{kas04} Kashikawa, N. et al. 2004, \pasj, 56, 1011
\bibitem[Kashikawa et al.(2006)]{kas06} Kashikawa, N. et al. 2006, \apj, 648, 7 (K06)
\bibitem[Kodaira et al.(2003)]{kod03} Kodaira, K. et al. 2003, \pasj, 55, L17
\bibitem[Kobayashi et al.(2007)]{kob07} Kobayashi, M.A.R., Totani, T., \& Nagashima, M. 2007, \apj, 670, 919
\bibitem[Kobayashi et al.(2010)]{kob10} Kobayashi, M.A.R., Totani, T., \& Nagashima, M. 2010, \apj, 708, 1119
\bibitem[Komatsu et al.(2009)]{kom09} Komatsu, E. et al. 2009, \apjs, 180, 330
\bibitem[Le Delliou et al.(2005)]{led05} Le Delliou, M., Lacey, C., Baugh, C.M., Guiderdoni, B., Bacon, R., Courtois, H., Sousbie, T., Morris, S.L. 2005, \mnras, 357, L11
\bibitem[Leitherer et al.(1995)]{lei95} Leitherer, C., Robert, C., \& Heckman, T.M. 1995, \apjs, 99, 173
\bibitem[Malkan et al.(2003)]{mal03} Malkan, M., Webb, W., \& Konopacky,Q. 2003, \apj, 598, 878
\bibitem[Maddau, Haardt, \& Rees(1999)]{mad99} Maddau, P., Haardt, F., \& Rees, M.J. 1999, \apj, 514, 648
\bibitem[Malhotra \& Rhoads(2004)]{mal04} Malhotra, S. \& Rhoads, J.E. 2004, \apj, 617, L5
\bibitem[Mas-Hesse et al.(2003)]{mas03} Mas-Hesse, J.M., Kunth, D., Tenorio-Tagle, G., Leitherer, C., Terlevich, R. J., Terlevich, E. 2003, \apj, 598, 858
\bibitem[McLinden et al.(2011)]{mcl11} McLinden, E.M. et al. 2011, \apj, 730, 136
\bibitem[McQuinn et al.(2007)]{mcq07} McQuinn, M. et al. 2007, \mnras, 381, 75
\bibitem[Mesinger \& Furlanetto (2007)]{mes07} Mesinger, A. \& Furlanetto, S. 2007, \apj, 669, 663
\bibitem[Nagamine et al.(2010)]{nag10} Nagamine, K. et al. 2010, \pasj, 62, 1455
\bibitem[Nagao et al.(2007)]{nag07} Nagao, T. et al. 2007, \aap, 468, 877
\bibitem[Nakamura et al.(2011)]{nak11} Nakamura, E. et al. 2011, \mnras, 412, 2579
\bibitem[Neufeld (1991)]{neu91} Neufeld, D.A. 1991, \apjl, 370, 85
\bibitem[Nilsson et al.(2009)]{nil09} Nilsson, K.K. et al. 2009, \mnras, 400, 232
\bibitem[Oesch et al. (2010)]{oes10} Oesch, P.A. et al. 2010, \apj, 709, L16
\bibitem[Ono et al.(2010)]{ono10} Ono, Y. et al. 2010, \apj, 724, 1524
\bibitem[Ota et al.(2008)]{ota08} Ota, K. et al. 2008, \apj, 677, 12
\bibitem[Ota et al.(2010)]{ota10} Ota, K. et al. 2010, \apj, 722, 803
\bibitem[Ouchi et al.(2008)]{ouc08} Ouchi, M. et al., 2008, \apjs, 176, 301
\bibitem[Ouchi et al.(2010)]{ouc10} Ouchi, M. et al., 2010, \apj, 723, 869
\bibitem[Pawlik et al.(2009)]{paw09} Pawlik, A. et al. 2009, \mnras, 394, 1812
\bibitem[Pentericci et al. (2010)]{pen10} Pentericci, L. et al. 2010, \aap, 514, 64
\bibitem[Razoumov \& Sommer-Larsen (2010)]{raz10} Razoumov, A.O., \& Sommer-Larsen, J. 2010, \apj, 710, 1239
\bibitem[Reddy et al.(2008)]{red08} Reddy, N.A. et al. 2008, \apjs, 175, 48
\bibitem[Richard et al. (2008)]{ric08} Richard, J. et al. 2008, \apj, 685, 705
\bibitem[Ricotti \& Shull(2000)]{ric00} Ricotti, M. \& Shull, J.M. 2000, \apj, 542, 548
\bibitem[Shapley et al.(2001)]{sha01} Shapley, A.E. et al. 2001, \apj, 562, 95
\bibitem[Shapley et al.(2003)]{sha03} Shapley, A.E., Steidel, C.C., Pettini, M., Adelberger, K.L. 2003, \apj, 588, 65
\bibitem[Shapley et al.(2006)]{sha06} Shapley, A.E. et al. 2006, \apj, 651, 688
\bibitem[Schaerer(2002)]{sch02} Schaerer, D. 2002, \aap, 382, 28
\bibitem[Shimasaku et al.(2006)]{shi06} Shimasaku, K. et al. 2006, \pasj, 58, 313 (S06)
\bibitem[Siana et al.(2010)]{sia10} Siana, B. et al. 2010, \apj, 723, 241
\bibitem[Somerville et al.(2004)]{som04} Somerville, R.S., Lee, K., Ferguson, H.C., Gardner, J.P., Moustakas, L.A., Giavalisco, M. 2004, \apj, 600, L171
\bibitem[Stanway et al.(2007)]{sta07} Stanway, E.R. et al. 2007, \mnras, 376, 727
\bibitem[Stark et al.(2007)]{sta07} Stark, D.P. et al. 2007, \apj, 663, 10
\bibitem[Stark et al.(2010)]{sta10} Stark, D.P. et al. 2010, \mnras, 408, 1628
\bibitem[Stark et al.(2011)]{sta11} Stark, D.P. et al. 2011, \apj, 728, L2
\bibitem[Taniguchi et al.(2010)]{tan10} Taniguchi, Y. et al. 2010, \apj, 724, 1480
\bibitem[Taniguchi et al.(2005)]{tan05} Taniguchi, Y. et al. 2005, \pasj, 57, 165
\bibitem[Tilvi et al.(2010)]{til10} Tilvi, V. et al. 2010, \apj, 721, 1853
\bibitem[Vanzella et al. (2010)]{van10} Vanzella, E. et al. 2010, \apj, 725, 1011
\bibitem[Vanzella et al. (2009)]{van09} Vanzella, E. et al. 2009, \apj, 695, 1163
\bibitem[Verhamme et al. (2008)]{ver08} Verhamme, A. et al. 2008, \aap, 491, 89
\bibitem[Westra et al.(2005)]{wes05} Westra, E., Jones, D.H., Lidman, C.E., Athreya, R.M., Meisenheimer, K., Wolf, C., Szeifert, T., Pompei, E., Vanzi, L. 2005, \aap, 430, L21
\bibitem[Willott et al.(2005)]{wil05} Willott, C.J et al., 2005, \apj, 633, 630
\bibitem[Wise \& Cen(2009)]{wis09} Wise, J.H., \& Cen, R. 2009, \apj, 693, 984
\bibitem[Wyithe \& Loeb(2004)]{wyi04} Wyithe, J.S.B., \& Loeb, A., 2004, \nat, 432, 194
\bibitem[Yoshida et al.(2007)]{yos07} Yoshida, N., et al. 2007, \apj, 663, 687
\bibitem[Zheng et al.(2010)]{zhe10} Zheng, Z., et al. 2010, \apj, 716, 574

\end{thebibliography}
\end{document}